\documentclass[11pt,aps,preprint,showpacs,amsmath,amssymb,amsfonts,eqnarray,array]{revtex4}
\usepackage{anysize}
\marginsize{1.9cm}{1.9cm}{1.55cm}{1.55cm}
\usepackage{graphicx}
\usepackage{dcolumn}
\usepackage{bm}
\begin{document}


\title{Inelastic electron and Raman scattering from the collective excitations in quantum wires:
Zero magnetic field\\}
\author{Manvir S. Kushwaha}
\affiliation
{\centerline {Department of Physics and Astronomy, Rice University, P.O. Box 1892, Houston, TX 77251, USA}}

\date{\today}

\begin{abstract}
The nanofabrication technology has taught us that an $m$-dimensional confining potential imposed upon an
$n$-dimensional electron gas paves the way to a quasi-($n$-$m$)-dimensional electron gas, with $m \le n$
and $1\le n, m \le 3$. This is the road to the (semiconducting) quasi-$n$ dimensional electron gas
systems we have been happily traversing on now for almost two decades. Achieving quasi-one dimensional
electron gas (Q-1DEG) [or quantum wire(s) for more practical purposes] led us to some mixed moments in
this journey: while the reduced phase space for the scattering led us believe in the route to the faster
electron devices, the proximity to the 1D systems left us in the dilemma of describing it as a Fermi
liquid or as a Luttinger liquid. No one had ever suspected the potential of the former, but it took
quite a while for some to convince the others on the latter. A realistic Q-1DEG system at the low
temperatures is best describable as a Fermi liquid rather than as a Luttinger liquid. In the language of
condensed matter physics, a critical scrutiny of Q-1DEG systems has provided us with a host of exotic
(electronic, optical, and transport) phenomena unseen in their higher- or lower-dimensional counterparts.
This has motivated us to undertake a systematic investigation of the inelastic electron scattering (IES)
and the inelastic light scattering (ILS) from the elementary electronic excitations in quantum wires. We
begin with the Kubo's correlation functions to derive the generalized dielectric function, the inverse
dielectric function, and the Dyson equation for the dynamic screened potential in the framework of
Bohm-Pines' random-phase approximation. These fundamental tools then lead us to develop methodically the
theory of IES and ILS for the Q-1DEG systems. As an application of the general formal results, which know
no bounds regarding the subband occupancy, we compute the density of states, the Fermi energy, the full
excitation spectrum [comprised of intrasubband and intersubband single-particle as well as collective
excitations], the loss functions for the IES, and the Raman intensity for the ILS. We observe
that it is the collective (plasmon) excitations that largely contribute to the predominant peaks in the
energy-loss and the Raman spectra. The inductive reasoning is that the IES can be a potential alternative
of the overused ILS for investigating collective excitations in quantum wires. We trust that this research
work shall be useful to all -- from novice to expert and from theorist to experimentalist -- who believe in
the power of traditional science.
\end{abstract}

\pacs{73.21.Hb; 74.25.nd; 78.67.Lt; 79.20.Uv}
\maketitle


\newpage

\section{Introduction}

Thanks to the advancements in the nanofabrication technology and the electron lithography, condensed matter
physics of the past two and a half decades is largely the physics of the non-conventional (i.e., the
lower-dimensional) quantum systems. These are the man-made, semiconducting, quasi-n-dimensional electron gas
(Q-nDEG) systems, with $n$ [$\equiv 2$, 1, or 0] being the directional degree(s) of freedom, in which the
charge carriers subjected to external probes such as electric and/or magnetic field can induce exotic quantal
effects that strongly modify their behavior characteristics. A Q-2DEG is known to serve as the basis for the
fabrication of still-lower dimensional electron systems (nDES) such as quantum wires (or Q-1DEG) and quantum
dots (or Q-0DEG). With the lowering of the system's dimensions from 3 to 2, 2 to 1, and 1 to 0, the density
of (electronic) states assumes, respectively, the shape of the stair case, the spikes, and the
$\delta$-function-like (discrete) peaks. Also, being recognized is the fact that the lower the dimension (or
the degree of freedom), the greater the role of the many-body effects on the electronic properties of the
system. A comprehensive review of the electronic, optical, and transport phenomena in the lower dimensional
systems can be found in Ref. 1.

The research pursuit behind the quantum wires, which lie in the middle of the quantum rainbow comprising the lower-dimensional systems, has (almost) always been motivated by the fact that the 1D k-space restriction
severely reduces the impurity scattering, thereby significantly improving upon the low-temperature electron
mobilities. Thus, the technological promise that breaks out is the path to the ultrafast, low-threshold
electronic devices synthesized out of the quantum wires [2]. The quantum wires [or, more realistically, a
Q-1DEG for better and broader range of physical understanding] have exhibited some unique transport properties
such as the magnetic depopulation, electron waveguide, quenching of the quantum Hall effect, quantization of
conductance, negative-energy dispersion with the magnetic field, magnetoroton excitations, spin-charge
separation, and, most recently, their role as the optical amplifiers [3]. The details of these phenomena have
been systematically discussed in Ref. 1.

It is noteworthy that the early theoretical advances in the quantum wires suffered from an intense debate
over whether the system is best describable as Luttinger liquid or as Fermi liquid [1]. The essential
feature that differentiates the Luttinger liquid (LL) from the normal Fermi liquid (FL) is that the LL,
unlike the FL, is devoid of the Fermi surface. This implies that the momentum distribution function $n_k$
is continuous through the Fermi momentum $k_F$. The LL picture can be attributed to three basic notions:
(i) the electron-phonon-coupling--induced lattice Peierls distortion, (ii) the disorder--induced Anderson
localization, and (iii) the electron-electron-interaction--induced strong correlation. However, it has
been consistently argued [4] and experimentally justified [5] that the {\em realistic} quantum wires are
nearly free from the Peierls distortion at lower temperatures, they are pure enough to avoid the Anderson
localization, and impure enough to inhibit the strongly correlated behavior. Morevoer, the long-wavelength
plasmon dispersion and the (Tomonaga-Luttinger) boson dispersion of the 1DES are shown to be {\em similar}
due to the vanishing of vertex corrections in the single-particle polarizability function of the 1DEG [6].

As such, it is fair to say that no other system seems to have engaged the researchers with as many appealing
features to explore. And yet, there are some interesting fundamental issues which have not drawn any
attention of the scientific community so far. One of them is, for sure, the systematic and rigorous theory
of the inelastic electron scattering from the elementary electronic excitations in the quantum wires. This is
an important void and filling this void is one of the main themes behind the present work. To be specific, we
embark on a systematic and rigorous mathematical machinery for the theory of inelastic electron scattering
(IES) and of inelastic light (or Raman) scattering (ILS) from the elementary electronic excitations in the
experiments on the quantum wires in the absence of an applied magnetic field. As to the electronic excitations,
we will focus on the (single-particle as well as the collective) charge-density excitations in the Q-1DEG in
the isolated quantum wires. The charge-density excitations (CDE) subsist through the direct Coulomb
interactions just as the spin-density excitations (SDE) subsist through the exchange-correlation Coulomb
interactions. The energy shift of the collective CDE (SDE) from the corresponding (bare) single-particle
transition energy is a direct measure of the many-body effects such as depolarization (excitonic) shift [1].

Despite the fact that they remained (regrettably) unnoticed, the first decisive steps in studying the electronic
excitations in organic structures mimicking the present-day quantum wires were taken -- theoretically [7-11] and experimentally [12-15] -- much before the actual era of semiconducting quantum wires began. Interestingly enough,
the first (direct) plasmon dispersion and damping in TTF-TCNQ were observed by Ritsko et al. [15], using
high-energy inelastic electron scattering in thin films at 300 K. The current theoretical trends for
investigating plasmon dispersion in the semiconducting quantum wires began later [16] followed by a host of
extensive works on the subject [17-27], using, often, similar variants of the self-consistent field
approximation [28]. On the practical side, a vast majority of experiments aimed at exploring the Q-1DEG were
motivated by the early prediction that quantum wires could pave the way to very high-mobility transistors due to
the reduced phase space for the carrier scattering. There is a long list of early experiments which mainly
focused on the fabrication of quantum wire structures and their characterization by demonstrating, e.g., the
quantization into 1D subbands and specifying the subband spacings, Fermi energy, carrier concentration, and
subband occupancy, ... etc. [see, e.g., Ref. 1]. The experiments broadly employed to observe plasmon dispersion
in quantum wires include the far-infrared (FIR) spectroscopy [29-35], photoluminescence (PL) spectroscopy [36-39],
and Raman scattering (RS) spectroscopy [40-45].

It was soon realized that the FIR experiments are unable to yield any trace of many-body effects whose observation
is restrained for these reasons: (i) due to the generalized Kohn theorem (GKT) [46] only the lowest intersubband
CDE couples to the FIR radiation for a bare harmonic potential, (ii) a different -- usually magnetotransport --
experiment needs to be carried out in order to compare the energies of the observed collective and single-particle
CDE, and (iii) different types of collective excitations involving changes in the spin density (SDE) cannot be
observed [1]. The SDE are of unusual interest, because, unlike the CDE, they are only influenced by the exchange
and correlation parts of the electron-electron interactions, and hence their observation allows a clear distinction
between direct and excitonic (or final state) interactions. These restrictions are overcome by performing instead
Raman scattering by electrons, which is well suited to observe not only the collective CDE and SDE, but also the
corresponding SPE, which remain unaffected by the many-body effects. The Raman spectroscopy makes all three
excitations identifiable by simple polarization selection rules [1].

The present paper aims at developing a comprehensive theory of inelastic electron scattering and inelastic light
(or Raman) scattering in the isolated quantum wires in the absence of an applied magnetic field. To this end, we
do require a systematic knowledge of the full excitation spectrum for the sake of justifying and identifying the
loss peaks in the IES and the intensity peaks in the ILS. For this purpose, we derive the nonlocal, dynamic
dielectric function, inverse dielectric function, Dyson equation, and other correlation functions within the
framework of Bohm-Pines' full random-phase approximation (RPA) [28]. Since it is the first paper of its kind
embarking on the theory of IES and ILS in the quantum wires, we keep it to the bare-bone simplicity and hence
neglect, for the time being, the many-particle (exchange-correlation) interactions. The complexity should (and
will) come later. It is noteworthy that the author had undertaken a similar endeavor of developing an extensive
theory of IES and ILS in quantum wells recently [47]. Therefore, we will considerably rely on Ref. 47
 -- whenever deemed appropriate -- in order to minimize the overlap.

As will be seen shortly, the probability (or loss) function for the IES is defined in terms of the inverse
dielectric function (IDF), whereas the Raman intensity (or the cross-section) for the ILS is expressed in terms
of the total (or interacting) density-density correlation function (DDCF). The latter essentially needs to
introduce the double-time (retarded) Green-function [48] whose equation of motion is solved by making rigorous
use of the rules of second quantization, the Fourier transforms, and the effective-mass approximation [49]. The
illustrative examples reveal the supremacy of the collective (plasmon) excitations to the predominant loss peaks
in the IES as well as the Raman peaks in the ILS.

The rest of the article is organized as follows. In Sec. II, we introduce the theoretical framework leading to
the derivation of the generalized nonlocal, dynamic, dielectric function, screened potential, the inverse
dielectric function, the Dyson equation, the probability (or loss) functions defining the inelastic electron
scattering, and the differential scattering cross-section for the inelastic light scattering in the isolated
quantum wires. We also offer a critical analytical diagnoses of some of the exact results, discuss their
relevant aspects, and specify them for the practicality in order to fully address the solution of the problem.
In Sec. III, we discuss several illustrative examples of, for instance, the excitation spectrum consisting of
the single-particle and collective (plasmon) excitations, the electron energy-loss spectra, the Raman intensity
spectra, and the inverse dielectric functions. Finally, in Sec. IV, we summarize our observations and propose
some distinctive features worth adding to the problem.

\begin{figure}[htbp]
\includegraphics*[width=8cm,height=9cm]{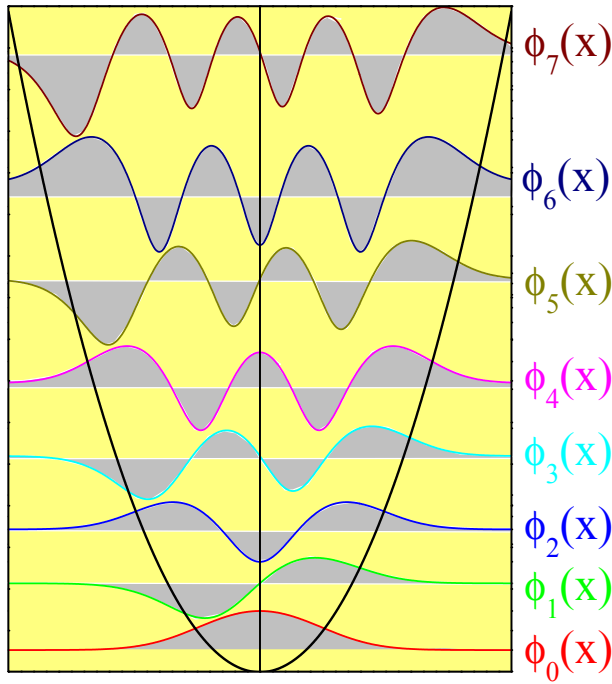}
\caption{(Color online) The 1D confining parabolic potential well: An ideal parabolic potential (such as this)
represents a {\em harmonic oscillator} whose eigenfunction and eigenenergy can be calculated analytically [see,
e.g., Eqs. (3) and (4)]. The eigenfunctions show an {\em even-odd} alternation just as in the case of a
symmetric, square quantum well.}
\label{fig1}
\end{figure}

\section{Methodological Framework}

\subsection{The eigenfunctions and eigenenergies}

It is now well-known that the high-resolution lithography can make possible the fabrication of semiconductor
structures in which the confinement of electronic motion in two dimensions paves the way to achieve a Q-1DEG.
An {\em extreme} confinement along the z direction restrains the electron motion to the lowest 2D subband and
hence allows us drop the z coordinate from further consideration. To be succinct, we start with a 2DEG in the
x-y plane with a confining harmonic potential along the x direction [see, e.g., Fig. 1]. The resulting system
is a typical Q-1DEG with a free electron motion along the y direction and the size quantization in the x
direction. For such a system, the single-particle Hamiltonian can be clearly expressed as
\begin{equation}
H =\frac{\hat{p}_y^2}{2\, m^*} \, + \, \frac{1}{2}\, m^* \,\omega^2_0 \, x^2\, ,
\end{equation}
where $\hat{p}_y=-i\,\hbar\nabla_{y}$ is the momentum operator component along the y axis and $\omega_0$ the
characteristic frequency of the harmonic oscillator. In this situation, the resultant Q-1DEG system is
characterized by the eigenfunction
\begin{equation}
\psi_j({\boldsymbol r})=\frac{1}{\sqrt{L_y}}\,e^{i\,k_y\, y} \,\phi_n(x)\, ,
\end{equation}
where ${\boldsymbol r} \equiv (x, y)$ is a 2D vector in the direct space [or r-space], $L_y$ the normalization
length, $j\equiv k,n$ the composite quantum index, and the Hermite function $\phi_n(x)$ is defined as
\begin{equation}
\phi_n(x)=N_n\,e^{-x^2/(2\ell^2_c)}\,H_n(x/\ell_c)\, ,
\end{equation}
where $n$, $N_n=(\sqrt{\pi}\,2^n\,n!\,\ell_c)^{-1/2}$, $\ell_c=\sqrt{\hbar/(m^*\omega_0)}$, and $H_n$($x$) are,
respectively, the subband index due only to the size quantization, the normalization constant, the characteristic
length of the harmonic oscillator, and the Hermite polynomial of order $n$, and the eigenenergy
\begin{equation}
\epsilon_n (k)= (n+\frac{1}{2})\,\hbar\,\omega_0 \,+\,\frac{1}{2m^*}\,\hbar^2\,k_{y}^2\, ,
\end{equation}
where ${k}\equiv \mid {\boldsymbol k}\mid$ and ${\boldsymbol k} \equiv (k_x, k_y)$ is a 2D wave vector in the
reciprocal space [or k-space]. The first term on the right-hand side refers to the energy of the $n$th
subband. Any standard textbook on quantum mechanics tells us that in 1D case each energy level corresponds to a
unique quantum state and hence the system as such stands as non-degenerate. This can quite easily be verified
from the degree of degeneracy $g_n=\binom {N+n-1}{n}$, where $N$ ($n$) stands for the dimension (quantum number).
The perceivable advantage of working with the harmonic potential is that one can go to a greater extent in order
to carry out complex analytical work to solve, for example, a 1D Schr\"odinger equation and deduce the exact
form of the eigenfunction and the eigenenergy. It turns out, however, that the general formalism constructed in
this paper is independent of any specific form of the confinement potential.

\subsection{The correlation functions: Design and derivation}

This section is devoted to design and derive the most important correlation functions useful for investigating
the collective excitations in quantum wires. These are the nonlocal, dynamic dielectric function, the screened
interaction potential, and the inverse dielectric function. While a few authors [16-21] have attempted to
calculate the dielectric function (differently), no one has, to the author's knowledge, ever derived the latter
two for the system of quantum wires. It is also fair to state that our derivation of these correlation functions
is, in general, noticeably comprehensive and diligent. As such, this section stands out to be more than just a review of the standard correlation functions.

\subsubsection{The nonlocal, dynamic dielectric function}

At the outset, it is worth mentioning that there is really no need to specify the material media so far as the
formalism building is concerned. Since $k_x$ will not appear until the very end, we choose to drop the subscript
y on all the quantities for the sake of brevity. It is always interesting, important, and convenient to begin
with the expression for the single-particle density-density correlation function (DDCF) given by [1]

\begin{equation}
\chi^{0} ({\boldsymbol r},{\boldsymbol r'}; \omega)=\sum_{ij}\, \Lambda_{ij}\,\,
\psi^*_i ({\boldsymbol r'})\,\psi_j ({\boldsymbol r'})\,
\psi^*_j ({\boldsymbol r})\,\psi_i ({\boldsymbol r})\, ,
\end{equation}
where the subindex $i,j\equiv k,n$ and the substitution $\Lambda_{ij}$ is defined by
\begin{equation}
\Lambda_{ij}= 2\, \frac{f(\epsilon_i)-f(\epsilon_j)}{\epsilon_i-\epsilon_j+\hbar\omega^+} \, ,
\end{equation}
Here $f(x)$ is the familiar Fermi distribution function. $\omega^+=\omega+i\gamma$ and small but nonzero $\gamma$
refers to the adiabatic switching of the Coulombic interactions in the remote past. The factor of $2$ accounts for
the spin degeneracy. Making use of Eq. (2) in Eq. (5) leaves us with
\begin{eqnarray}
\chi^{0}(x, y; x', y'; \omega)
=\frac{1}{L^2}\sum_{nn'}\,\sum_{kk'}&&\Lambda_{nn'}(k, k';\omega)\,e^{-i\,q\,(y-y')}\nonumber\\
              && \times \, \phi^*_{n}(x')\,\phi_{n'}(x')\,\phi^*_{n'}(x)\,\phi_{n}(x)\, ,
\end{eqnarray}
where $q=k'-k$ is the momentum transfer. Since there is absolutely no loss of translational invariance along the
$y$ axis, one can always Fourier transform both sides with respect to $y$. As such, multiplying both sides of this
equation by $e^{iq'(y-y')}$ and integrating with respect to $y$ yields
\begin{eqnarray}
\chi^0(x, x'; q',\omega)
&=&\frac{2\pi}{L^2}\,\sum_{nn'}\,\sum_{kk'}\,\Lambda_{nn'}(k, k';\omega)\, \delta(q-q')\nonumber\\
&& \hspace{2.2cm}\times \, \phi^*_{n}(x')\,\phi_{n'}(x')\,\phi^*_{n'}(x)\,\phi_{n}(x)\nonumber\\
&=&\frac{1}{L}\,\sum_{nn'}\,\sum_{k}\,\Lambda_{nn'}(k, k'=k+q';\omega)\, \nonumber\\
&& \hspace{2.2cm}\times \, \phi^*_{n}(x')\,\phi_{n'}(x')\,\phi^*_{n'}(x)\,\phi_{n}(x)\,
\end{eqnarray}
The second equality is obtained by opening up the sum over $k'$ in terms of an integral and solving that integral
with the help of the Dirac $\delta$ function. Since $q'$ is a dummy variable, one can simply ignore the prime to
write this equation as follows.
\begin{eqnarray}
\chi^0(x, x'; q, \omega)
&=&\frac{1}{L}\,\sum_{nn'}\,\sum_{k}\,\Lambda_{nn'}(k, k'=k+q;\omega)\, \nonumber\\
&&  \hspace{2.2cm}  \times \, \phi^*_{n}(x')\,\phi_{n'}(x')\,\phi^*_{n'}(x)\,\phi_{n}(x)\nonumber\\
&=& \sum_{nn'}\,\Pi_{nn'}(q, \omega)\,\phi^*_{n}(x')\,\phi_{n'}(x')\,\phi^*_{n'}(x)\,\phi_{n}(x)\, ,              \end{eqnarray}
where
\begin{equation}
\Pi_{nn'}(q, \omega)
=\frac{1}{L}\,\sum_k \,\Lambda_{nn'}(...)
=\frac{2}{L}\,\sum_k \,\frac{f(\epsilon_{nk})-f(\epsilon_{n'k'})} {\epsilon_{nk}-\epsilon_{n'k'}+\hbar\omega^+}
\end{equation}
is, generally, termed as the polarizability function. Next, let us make use of the Kubo's correlation function
in order to write the induced particle density expressed as
\begin{align}
n_{in}(q, \omega; x)
&= \int dx'\, \chi(q, \omega; x, x')\,V_{ex}(q, \omega; x') \nonumber\\
&= \int dx'\, \chi^0(q, \omega; x, x')\,V(q, \omega; x')
\end{align}
Here $V(...)=V_{ex}(...)+V_{in}(...)$ is the total potential, with subscripts ex (in) referring to the external
(induced) potential. Let us mention {\em once and for all} that although we use the term {\em potential}, we
always mean them to be the {\em potential energy}. Here $\chi (...)$ [$\chi^0$(...)] is the total [single-particle]
DDCF. Eqs. (11) are actually already Fourier-transformed with respect to the spatial coordinate y and the time $t$.
Moreover, the induced potential in terms of the induced particle density is given by
\begin{equation}
V_{in}(q, \omega; x)=\int dx'\, V_{ee}(q; x, x')\, n_{in}(q, \omega; x')
\end{equation}
where $V_{ee}(...)$ is the 1D Fourier transform of the binary Coulomb interactions and is defined by
\begin{equation}
V_{ee}(q; x, x')=\frac{2e^2}{\epsilon_b}\, K_0(q\mid x-x'\mid)
\end{equation}
where $K_0 (x)$ is the zeroeth-order modified Bessel function of the second kind, which diverges as $-\ln (x)$
when $x\rightarrow 0$. Here $\epsilon_b$ is the background dielectric constant of the medium the Q-1DEG is
embedded in. Notice, before we proceed further, that it is quite straightforward to prove, from Eqs. (11) and
(12), that $\chi(...)$ and $\chi^0(...)$ are actually correlated through the famous Dyson equation [see Fig. 2]
\begin{equation}
\chi (x, x')=\chi^0 (x, x') +\int dx'' \int dx''' \, \chi^0 (x, x'')\, V_{ee}(x'', x''')\, \chi (x''', x')\, ,
\end{equation}
where the ($q, \omega$)-dependence is suppressed for the sake of brevity. Equation (12), with the aid of Eqs. (9)
and (11), yields
\begin{align}
V_{in}(q, \omega; x)
&=\int dx'\, \int dx''\, V_{ee}(q; x, x')\,\chi^0(q, \omega; x', x'')\,V(q, \omega;x'')\nonumber\\
&=\frac{2\,e^2}{\epsilon_b}\,\sum_{nn'}\,\Pi_{nn'}(q, \omega)\,
        \int dx' \int dx''\, K_0(q\mid x-x'\mid)\, V(q, \omega; x'')\nonumber\\
& \hspace{5.0cm} \times\,\phi^*_{n}(x'')\,\phi_{n'}(x'')\,\phi^*_{n'}(x')\,\phi_{n}(x').
\end{align}
Let us now take the matrix elements of both sides of Eq. (15) between the states $\mid m \left. \right >$ and
$\mid m' \left. \right > $ to write
\begin{equation}
\left < m  \mid V_{in}(...)\mid m' \right >=\sum_{nn'}\, \Pi_{nn'}(q, \omega)\,F_{mm'nn'}(q)\,
\left < n \mid V(...)\mid n' \right >\, ,
\end{equation}
where the substitution $F_{mm'nn'}(q)$ defined by
\begin{equation}
F_{mm'nn'}(q)=\frac{2\,e^2}{\epsilon_b}\, \int dx \int dx' \,
\phi^*_{m}(x)\,\phi_{m'}(x)\, K_0(q\mid x-x'\mid)\,\phi^*_{n'}(x')\,\phi_{n}(x')\, ,
\end{equation}
refers to the matrix elements of the Coulombic interactions. Employing the condition of self-consistency --
$V(...)=V_{ex}(...)+V_{in}(...)$ -- leads us to cast Eq. (16) in the form
\begin{equation}
\left < m  \mid V_{ex}(...)\mid m'\right > =\sum_{nn'}\,
\left [\delta_{mn}\delta_{m'n'} - \Pi_{nn'}(q, \omega)\,F_{mm'nn'}(q) \right ]\,
\left < n \mid V(...)\mid n' \right >\, ,
\end{equation}
Now, since $V_{ex}(...)$ and $V(...)$ are related via
\begin{equation}
\begin{split}
V_{ex}(y)&=\int dx' \,\epsilon (x, x')\, V(x')\, , \qquad{\rm or}\\
\left < m \mid V_{ex}(...)\mid m' \right > &=
\sum_{nn'}\,\epsilon_{mm'nn'}(q, \omega)\, \left < n \mid V(...) \mid n' \right >\, ,
\end{split}
\end{equation}
comparing Eqs. (18) and (19) finally yields the generalized nonlocal dynamic dielectric function expressed by
\begin{equation}
\epsilon_{mm'nn'}(q, \omega)=\delta_{mn}\delta_{m'n'} - \Pi_{nn'}(q, \omega)\,F_{mm'nn'}(q)\, .
\end{equation}
This is clearly an outcome of the condition of self-sustaining the plasma oscillations in the electron density
when the external potential $V_{ex}$($\cdots$)$=0$. This implies that the excitation spectrum -- composed
of the single-particle as well as collective excitations -- should be computed through searching the zeros of the
determinant of the matrix $\tilde{\epsilon}$ ($q, \omega$) [i.e., $\mid \tilde{\epsilon}(q, \omega) \mid =0$].
This also demands limiting the number of subbands to be involved in the problem; otherwise, the general matrix,
with elements $\epsilon_{mm'nn'}(...)$, is an $\infty \times \infty$ matrix and hence impossible to solve. If we
are allowed to remind the reader, the situation is very much similar to the computation of the band structure in
the solid state physics: in order to truncate an $\infty \times \infty$ matrix obtained from the relevant secular
equations, we have to limit, e.g., the number of plane waves [$N=(2n+1)^2$; $n$ being an integer], which also
happens to define the dimension of the resulting matrix [$N \times N$] and hence the number of eigenvalues and
eigenstates.

\begin{figure}[htbp]
\includegraphics*[width=8cm,height=4.5cm]{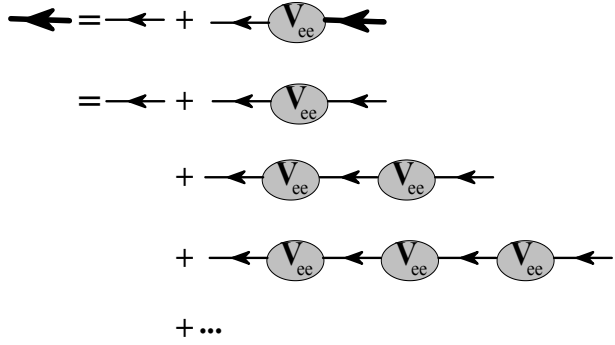}
\caption{A Feynman diagram for the Dyson equation: the thick [thin] line represents the reducible [irreducible]
DDCF $\chi (...)$ [$\chi^0 (...)$] in the full RPA. Here $V_{ee}$ is the 1D Fourier transform of the binary
Coulomb interactions. The arrows indicate the transition from initial to final spatio-temporal position
of the particle in the process.}
\label{fig2}
\end{figure}

\subsubsection{The inverse dielectric function}

A systematic knowledge of the inverse dielectric function (IDF) is of paramount importance not simply for its
own sake but also for its utility in studying, for instance, the inelastic electron scattering as well as the
inelastic light (or Raman) scattering in a given system [50]. Its role in the former (latter) is seen to be
explicit (implicit). Here, we derive the IDF methodically for the Q-1DEG in the absence of an applied magnetic
field. To start with, we first cast Eq. (15) in the form
\begin{align}
V_{ex}(x)
=\int dx' \Big [\delta(x-x')-\sum_{nn'}\,\Pi_{nn'}\,
         \int dx''\, \phi_{n}(x'')\,V_{ee}(x-x'')\,\phi^*_{n'}(x'') \nonumber\\
\times \, \phi^*_{n}(x')\,\phi_{n'}(x')\Big ]\,V(x') \, ,
\end{align}
again, suppressing the ($q, \omega$) dependence for brevity. Comparing this equation with Eq. (19) gives
\begin{align}
\epsilon(x, x')=\delta (x-x') -\sum_{nn'}\,\Pi_{nn'}\,
    \Big [\int dx'' \, \phi_n(x'')\,V_{ee}(x, x'')\,\phi^*_{n'}(x'')\Big ]
                    \phi^*_{n}(x')\,\phi_{n'}(x')\, .
\end{align}
This is the dielectric function we will determine the inverse of in this section. For this purpose, we first rewrite
Eq. (22) in the following way.
\begin{equation}
\epsilon(x, x')=\delta (x-x') -\sum_{\mu}\,L^*_{\mu}(x)\,\Pi_{\mu}\,S_{\mu}(x')\, .
\end{equation}
Notice that this range of summation covers an interval ($a, b$), which can safely be taken to be ($\infty, \infty$)
for the generality. Here $\mu \equiv \{n n'\}$ is a composite index such that $\mu\equiv \mu_s, \mu_a$; with
subscript s (a) referring to the symmetric (asymmetric) function depending upon whether $n+n'$ is an odd or an even
number. The aforesaid scheme is quite general and only distinguishes the symmetric structures from the antisymmetric
ones. The substitutions $L_{\mu}(x)$ and $S_{\mu}(x)$ representing, respectively, the long-range and the short-range
parts of the response function are used for the convenience and are defined by
\begin{align}
L_{\mu}(x) & \equiv L_{nn'}(x)=\int dx'\, \phi^*_n(x')\,V_{ee}(x, x')\,\phi_{n'}(x')\, , \\
S_{\mu}(x) & \equiv S_{nn'}(x)=\phi^*_{n}(x')\,\phi_{n'}(x')\, .
\end{align}
Next, we presume the inverse dielectric function $\epsilon^{-1}(x, x')$ to be given by, say,
\begin{equation}
\epsilon^{-1}(x, x')=\delta (x-x') +\sum_{\nu}\,A^*_{\nu}(x)\,H_{\nu}\,B_{\nu}(x')\, ,
\end{equation}
such that the integral equation
\begin{equation}
\int dx'' \, \epsilon^{-1}(x, x'')\,\epsilon(x'', x')=\delta (x-x')
\end{equation}
is satisfied. Equation (27), with the aid of  Eqs. (23) and (26), becomes
\begin{eqnarray}
\sum_{\mu}\,L^*_{\mu}(x)\,\Pi_{\mu}\,S_{\mu}(x') &-&
\sum_{\nu}\,A^*_{\nu}(x)\,H_{\nu}\,B_{\nu}(x')\nonumber\\
&+& \sum_{\mu\nu}\,A^*_{\nu}(x)\,H_{\nu}\,\Pi_{\mu}\,S_{\mu}(x')\alpha_{\mu\nu}=0\, ,
\end{eqnarray}
where
\begin{equation}
\alpha_{\mu\nu}=\int dx\, L^*_{\mu}(x)\,B_{\nu}(x)\, .
\end{equation}
Notice that we can replace the subindex $\nu$ with $\mu$ in the second term on the left-hand side of Eq. (28) with
no loss of generality. As a result, Eq. (28) assumes the form
\begin{equation}
\sum_{\mu}\,L^*_{\mu}(x)\,\Pi_{\mu}\,S_{\mu}(x')
+ \sum_{\mu\nu}\,A^*_{\nu}(x)\,H_{\nu}\,
[\Pi_{\mu}\,S_{\mu}(x')\alpha_{\mu\nu} - B_{\nu}(x')\,\delta_{\mu\nu}]=0\, .
\end{equation}
Multiplying this equation with $L^*_{\gamma}(x')$ and then integrating with respect to $x'$ leaves us with
\begin{equation}
\sum_{\mu}\,L^*_{\mu}(x)\,\Pi_{\mu}\,\beta_{\gamma\mu}
+ \sum_{\mu\nu}\,A^*_{\nu}(x)\,H_{\nu}\,
[\Pi_{\mu}\,\beta_{\gamma\mu}\alpha_{\mu\nu} - \alpha_{\gamma\nu}\,\delta_{\mu\nu}]=0\, ,
\end{equation}
where the new substitution $\beta_{\gamma\mu}$ is defined as
\begin{equation}
\beta_{\gamma\mu}=\int dx\, L^*_{\gamma}(x)\,S_{\mu}(x)\, .
\end{equation}
Now, defining $B_{\mu}(x)$ as follows
\begin{equation}
B_{\mu}(x)=\lambda_{\mu}\,S_{\mu}(x)\, ,
\end{equation}
redefines $\alpha_{\mu\nu}$ in Eq. (29) such that
\begin{equation}
\alpha_{\mu\nu}=\lambda_{\nu}\,\beta_{\mu\nu}\, .
\end{equation}
Consequently, Eq. (31) yields
\begin{equation}
\sum_{\mu}\,\Big [L^*_{\mu}(x)\,\Pi_{\mu}
+ \sum_{\nu}\,A^*_{\nu}(x)\,H_{\nu}\,\lambda_{\nu}\,
(\Pi_{\mu}\,\beta_{\mu\nu} - \delta_{\mu\nu})\Big ]\,\Big [\beta_{\gamma\mu}\Big ]=0\, .
\end{equation}
Either the first or the second factor is zero. Since the second factor $\beta_{\gamma\mu} \ne 0$, we obtain
\begin{equation}
L^*_{\mu}(x)\,\Pi_{\mu}
= \sum_{\nu}\,A^*_{\nu}(x)\,H_{\nu}\,\lambda_{\nu}\,(\delta_{\mu\nu}-\Pi_{\mu}\,\beta_{\mu\nu})
\end{equation}
Assuming $\Lambda_{\mu\nu}$ to be the inverse of $(\delta_{\mu\nu}-\Pi_{\mu}\,\beta_{\mu\nu})$ such that
\begin{equation}
\sum_{\mu}\,\Lambda_{\gamma\mu}\,(\delta_{\mu\nu}-\Pi_{\mu}\,\beta_{\mu\nu})=\delta_{\gamma\nu}\, ,
\end{equation}
multiplying Eq. (36) by $\Lambda_{\gamma\mu}$, and summing over $\mu$ yields
\begin{eqnarray}
\sum_{\mu}\,L^*_{\mu}(x)\,\Pi_{\mu}\,\Lambda_{\gamma\mu}
&=& \sum_{\nu}\,A^*_{\nu}(x)\,H_{\nu}\,\lambda_{\nu}\delta_{\gamma\nu}\nonumber\\
&=& A^*_{\gamma}(x)\,H_{\gamma}\,\lambda_{\gamma}
\end{eqnarray}
The second equality then implies that
\begin{equation}
A^*_{\gamma}(x) = \frac{1}{H_{\gamma}\,\lambda_{\gamma}}\,
            \sum_{\mu}\,L^*_{\mu}(x)\,\Pi_{\mu}\,\Lambda_{\gamma\mu}\, .
\end{equation}
Thus, equation (26), with the aid of (33) and (39), is finally expressed in the form
\begin{equation}
\epsilon^{-1}(x, x')=\delta (x-x') + \sum_{\mu\nu}\,L^*_{\mu}(x)\,\Pi_{\mu}\,\Lambda_{\nu\mu}\,S_{\nu}(x')\, .
\end{equation}
{\em Whether this is exactly the correct inverse of $\epsilon (x, x')$ can easily be warranted by substituting
$\epsilon (x, x')$ from Eq. (23) and $\epsilon^{-1} (x, x')$ from Eq. (40) in Eq. (27)}. Clearly, this is not
the end of the story. We are still left with an important question to be addressed:
{\em Is $\epsilon^{-1}(x, x')$ in Eq. (40) the unique inverse of $\epsilon (x, x')$ in Eq. (23)}? In order to
make dead sure, let us answer this question in negation and suppose that
$\kappa^{-1} (x, x')$ [$\ne \epsilon^{-1} (x, x')$] is an another inverse of $\epsilon (x, x')$. This then
implies that there must be a matrix, say, $\Lambda'_{\mu\nu}$ satisfying an identity exactly similar to the
one given in Eq. (37), i.e.
\begin{equation}
\sum_{\mu}\,\Lambda'_{\gamma\mu}\,(\delta_{\mu\nu}-\Pi_{\mu}\,\beta_{\mu\nu})=\delta_{\gamma\nu}\,
\end{equation}
Subtracting Eq. (41) from Eq. (37) piecewise leaves us with
\begin{equation}
\sum_{\mu}\,(\Lambda_{\gamma\mu}-\Lambda'_{\gamma\mu})\,(\delta_{\mu\nu}-\Pi_{\mu}\,\beta_{\mu\nu})=0
\end{equation}
Clearly, the second term is {\em not} equal to zero; otherwise the identity in Eq. (37) or Eq. (41) makes no
sense. Therefore, the first term equated to zero is the only viable option in Eq. (42), i.e., $\Lambda_{\gamma\mu}=\Lambda'_{\gamma\mu}$. This leads us to deduce that
$\kappa^{-1} (x, x') = \epsilon^{-1} (x, x')$. We can thus infer that $\epsilon^{-1} (x, x')$ in Eq. (40) is
truly the correct and the unique inverse of $\epsilon (x, x')$ in Eq. (23).

\subsubsection{The screened interaction potential}

The screening of the electron-electron (e-e) interactions is an intrinsic part of the many-body problem in
condensed matter physics. Naively defined, screening is the damping of electric fields caused by the presence
of mobile charge carriers (electrons) in metals and semiconductors. Standard textbooks on electromagnetism
teach us that the Coulomb force between a pair of particles diminishes with distance as $r^{-2}$, whereas the
average number of particles at each distance $r$ is proportional to $r^2$, assuming the medium is fairly
isotropic. As a result, the charge fluctuation at any one point has non-negligible effects at large distances.
In real systems, the screening aspect is rather more complex than describable within the classic Thomas-Fermi
approximation (TFA). The two characteristic features of TFA are: (i) it is only valid when electron density
is very high, and (ii) it assumes that the electrons can respond at any given wave-vector. However, it is
(almost) impossible for an electron on or within the Fermi surface to respond at wave-vectors shorter than the
Fermi wave vectors. In mathematics (physics), this is related to the Gibbs phenomenon (Friedel oscillations)
[1].

In what follows, we discuss fully the nonlocal, dynamic free-carrier screening effects in terms of the screened
interaction potential related to the bare Coulomb potential in the quantum wires. To that end, we consider two
test electrons located at $\boldsymbol r$ and $\boldsymbol r'$ in the plane. Their interaction energy is given
in terms of the screened Coulomb potential defined by
\begin{equation}
V_s({\boldsymbol r}, {\boldsymbol r}'; t-t')=\int d{\boldsymbol r}''\,
\epsilon^{-1}({\boldsymbol r}, {\boldsymbol r}''; t-t')\, V_{ee}({\boldsymbol r}'', {\boldsymbol r}')\, ,
\end{equation}
where the binary Coulomb interaction $V_{ee}(...)$ in the direct space is defined by
\begin{equation}
V_{ee}({\boldsymbol r}, {\boldsymbol r'})
=\frac{e^2}{\epsilon_b}\,\frac{1}{\mid {\boldsymbol r}-{\boldsymbol r}'\mid}
=\frac{e^2}{\epsilon_b}\,\frac{1}{\mid (x-x')^2+(y-y')^2 \mid^{1/2}}\, ,
\end{equation}
Remember, Eq. (13) represents a 1D Fourier transform of this equation. Next, we take the Laplacian
[$\nabla^2_{{\boldsymbol r}'}$] of Eq. (43) and use the identity
\begin{equation}
\nabla^2\,V_{ee}({\boldsymbol r}, {\boldsymbol r}')=-4\pi\,e^2\,\delta({\boldsymbol r}-{\boldsymbol r}')\, ,
\end{equation}
to write
\begin{equation}
\nabla^2\,V_s({\boldsymbol r}, {\boldsymbol r}'; t-t')=-4\pi\,e^2\,
              \epsilon^{-1}({\boldsymbol r}, {\boldsymbol r}'; t-t')
\end{equation}
This clearly makes it obvious that the determination of the screened Coulomb potential is equivalent to the
determination of the inverse dielectric function [1]. In what follows in this section, we will henceforth
suppress the $(q; \omega)$ dependence for the sake of brevity. First, we rewrite Eq. (22), with the aid of
Eq. (9), in the form
\begin{equation}
\epsilon(x, x')=\delta (x-x') - \int dx'' \, V_{ee}(x, x'')\,\chi^0(x'',x')\, .
\end{equation}
Substituting this in the identity defined in Eq. (27) yields
\begin{align}
\epsilon^{-1}(x, x')
&=\delta(x-x') + \int dx''\int dx'''\, \epsilon^{-1}(x, x'')\,V_{ee}(x'', x''')\,\chi^0(x''', x')\nonumber\\
&=\delta(x-x') + \int dx''\,V_s(x, x'')\,\chi^0(x'', x')
\end{align}
where $V_s(x, x')$ defined by
\begin{equation}
V_s(x, x')=\int dx''\, \epsilon^{-1}(x, x'')\,V_{ee}(x'', x')
\end{equation}
now stands for the Fourier transform of Eq. (43) with respect to the spatial coordinate $y$ and the time.
Finally, substitution of Eq. (48) in Eq. (49) yields, after rearranging the terms,
\begin{equation}
V_s(x, x')=V_{ee}(x, x') + \int dx'' \int dx'''\, V_{ee}(x,x'')\,\chi^0(x'', x''')\,V_s(x''', x')
\end{equation}
This is the desired Dyson equation that correlates the screened interaction potential $V_s (x, x')$ to the
bare Coulombic potential $V_{ee} (x, x')$ via single-particle DDCF $\chi^0(z, z')$. Notice that Eq. (50) can
also be derived using Feynman diagrams [see Fig. 2] if we identify the thick (thin) line with arrow referring
to the screened (bare Coulomb) potential and $V_{ee}$ is replaced with $\chi^0 (x, x')$.

\subsection{Inelastic electron and light scattering: Novel approach and results}

Irrespective of their practicality and differences, inelastic electron scattering and inelastic light scattering experiments have indisputably become the \emph{part and parcel} of the condensed matter physics. They are being
employed to investigate electronic, optical, and transport phenomena of as well as to characterize the materials
of all shapes, sizes, and dimensions. This motivated the author to devise a thoughtful, systematic theoretical strategy to interpret the experimental observations of the collective excitations in quantum wires through both
of these electron and optical spectroscopies. In that sense, this section stands out for entirely original
research and novel results that enable us to make theoretical predictions for and/or confirm practical
observations of the experimental techniques.

\subsubsection{The inelastic electron scattering}

Historically, Fermi is known to have laid the conceptual foundation of the inelastic electron scattering [51]
-- a subject that has lately become (practically) known as the electron energy-loss spectroscopy (EELS).
Subsequently, Kramers adopted a similar approach to determine the stopping power due to conduction electrons
[52]. A decade later, Ritchie gave a necessary impetus to the concept of fast-particle energy loss to the
plasmons in thin films in a classic work [53]. Reviewing the early theoretical strategies and recalling the
the {\em science} behind the technology of spectrometers [54], the author has recently developed a
comprehensive dielectric response theory (DRT) of inelastic electron scattering in quantum wells (or Q-2DEG)
[47]. Here we design and develop the DRT to be applicable to the inelastic electron scattering in the quantum
wires (or Q-1DEG) in the absence of an applied magnetic field.
For the fact that the energy losses involved in the quantum wires are small -- on the order of a few meV -- we
are fully confident in its applicability to the quantum wires.

The mathematical machinery of the DRT adopted to that end is comprised of two basic ingredients: (i) the incoming
(coherent) electron beam is considered to be a classical trajectory, and (ii) the collective (plasmon) excitations
are described in a quantal manner within the RPA. Note that ${\bf r}$ will be treated as a 3D vector in the direct
space (or r$-$space) until and unless stated otherwise. The fast-particle with a charge distribution
$\rho({\boldsymbol r}, t)=-e\delta ({\boldsymbol r}-{\boldsymbol r}(t))$ enforces a Coulomb potential
\begin{equation}
V_{ex}({\boldsymbol r}, t)=-e\,\phi_{ex}({\boldsymbol r}, t)=\frac{e^2}
  {\mid {\boldsymbol r}- {\boldsymbol r}(t)\mid} \, .
\end{equation}
By taking its Laplacian we obtain
\begin{equation}
\nabla^2\,V_{ex}({\boldsymbol r}, t)=-4\,\pi\,e^2\,\delta({\boldsymbol r}- {\boldsymbol r}(t))\, .
\end{equation}
Remember, the problem is addressed in terms of an effective potential
$V ({\boldsymbol r}, t)=V_{ex}({\boldsymbol r}, t) + V_{in}({\boldsymbol r}, t)$. Next, we define the induced
particle density in terms of an induced potential such that
\begin{equation}
n_{in}({\boldsymbol r}, t)=-\frac{1}{4\,\pi\,e^2}\,\nabla^2[V ({\boldsymbol r}, t)-V_{ex}({\boldsymbol r}, t)]\, .
\end{equation}
Note that the low-case (big-case) $v$ denotes the velocity (potential). The dynamics of the classical trajectory is
defined as follows. The (coherent) particle beam polarizes the plasma within the Q-1DEG and the induced polarization
produces an electric field which exerts a force back on the particle approaching the system. The intent is to
calculate the total work done by the induced force in order to obtain the total energy loss suffered by the particle.
A proper decomposition of the resultant expression determines the energy distribution of those electrons which suffer
an inelastic scattering. The net effect is expressed in terms of the rate of energy loss
\begin{equation}
\frac{dW}{dt}=-{\boldsymbol v}(t)\cdot {\boldsymbol F}(t)\, ,
\end{equation}
where $t$, ${\boldsymbol v}$, and ${\boldsymbol F}$ are, respectively, the time, the velocity, and the induced force.
Since $t$ runs from $-\infty$ to $+\infty$, the particle comes to finish its specular trajectory with its energy loss
\begin{equation}
W=-Re \left [\int^{+\infty}_{-\infty}dt \, {\boldsymbol v}(t)\cdot {\boldsymbol F}(t) \right ]\, .
\end{equation}
When the total energy lost by the particle is defined by
\begin{equation}
W=\int d{q} \int d\omega \, \hbar\omega\, P({q}, \omega)\, ,
\end{equation}
where the quantity $P({q}, \omega)\, d{q}\, d\omega$ is termed as the probability that the incoming particle is
inelastically scattered into the range of energy losses between $\hbar\omega$ and $\hbar(\omega+d\omega)$, and
into the range of momentum losses between $\hbar{q}$ and $\hbar({q} + d{q})$. The angular resolved loss function
$P(\omega)$ fully specifies the kinematics of the fast particle at the detector. Next, we determine the induced
force defined by
\begin{equation}
{\boldsymbol F}=\int d{\boldsymbol r}\, n_{in}({\boldsymbol r}, t)\,\nabla V ({\boldsymbol r}, t)\,.
\end{equation}
This together with Eq. (53) and after carefully solving the involved vectorial relations [see, e.g., Ref. 47] yields
\begin{equation}
{\boldsymbol F}=-\nabla V ({\boldsymbol r}, t) {\big |}_{{\boldsymbol r}={\boldsymbol r}(t)}\, .
\end{equation}
Exploiting the translational invariance along the $y$ axis, all quantities can legitimately be Fourier transformed
with respect to the coordinate $y$ and the time $t$. As such, Fourier transforming Eq. (52) with respect to $y$ and
$t$ gives
\begin{equation}
\big (\frac{d^2}{d{\boldsymbol r}^2_{\bot}} -q^2 \big )\,V_{ex}(q,\omega; {\boldsymbol r}_{\bot})
=-4 \pi e^2\, \int dt\, e^{i[\omega t -q y(t)]}\,\delta ({\boldsymbol r}_{\bot}-{\boldsymbol r}_{\bot}(t))\, ,
\end{equation}
where ${\boldsymbol r}_{\bot}\equiv (x, z)$. Let us express the Fourier transform of the electrostatic Green function
$G (q; {\boldsymbol r}_{\bot}, {\boldsymbol r}'_{\bot})$ to write
\begin{align}
G (q; {\boldsymbol r}_{\bot}, {\boldsymbol r}'_{\bot})
&=\frac{1}{(2\pi)^2}\,\int d{\boldsymbol q}_{_{\bot}}\, G(q, {\boldsymbol q}_{_{\bot}})\,
e^{i{\boldsymbol q}_{_{\bot}}\cdot ({\boldsymbol r}_{\bot}-{\boldsymbol r}'_{\bot})}\nonumber\\
&=\frac{1}{(2\pi)^2}\,\int d{\boldsymbol q}_{_{\bot}}\,
\frac{e^{i{\boldsymbol q}_{_{\bot}}\cdot ({\boldsymbol r}_{\bot}-{\boldsymbol r}'_{\bot})}}
{{\boldsymbol q}^2_{_{\bot}}+q^2}\nonumber\\
&=\frac{1}{2\pi}\, K_0(q\mid {\boldsymbol r}_{\bot}-{\boldsymbol r}'_{\bot} \mid)\, ,
\end{align}
where ${\boldsymbol q}_{_{\bot}}\equiv (q_x, q_z)$ is a 2D vector in the k-space. With the aid of Eq. (60), Eq. (59)
is solved to write
\begin{align}
V_{ex}(q,\omega; {\boldsymbol r}_{\bot})
&=4 \pi e^2\,\int dt \int d{\boldsymbol r}'_{\bot}\,
   G (q; {\boldsymbol r}_{\bot}, {\boldsymbol r}'_{\bot})\,e^{i[\omega t -q y(t)]}\,
        \delta ({\boldsymbol r}'_{\bot}-{\boldsymbol r}_{\bot}(t)) \nonumber\\
&=2\, e^2\,\int dt\, e^{i[\omega t -q y(t)]}\, \int d{\boldsymbol r}'_{\bot}\,
     K_0(q\mid {\boldsymbol r}_{\bot}-{\boldsymbol r}'_{\bot} \mid)\,
              \delta ({\boldsymbol r}'_{\bot}-{\boldsymbol r}_{\bot}(t))\nonumber\\
&=2\, e^2\, \int dt \, e^{i[\omega t -q y(t)]}\,
                    K_0(q\mid {\boldsymbol r}_{\bot}-{\boldsymbol r}_{\bot}(t) \mid)\, .
\end{align}
Notice that we will henceforth confine ourselves to work within the 2D [or ($x, y$)] plane in order to make a good
amalgam of the quantities involved. This implies that ${\boldsymbol r}_{\bot}$ in Eq. (61) only needs to be
replaced with $x$. Assuming the weak perturbation (and small energy losses), the particle beam is deemed to be
moving with a uniform velocity and its trajectory described by
\begin{equation}
{\boldsymbol r}\,(t)=x(t)\, \hat{x} + y(t)\, \hat{y}=(v_x t + x_0)\,\hat{x} + (v_y t + y_0)\, \hat{y}\, ,
\end{equation}
where ${\boldsymbol r}$ [$\equiv$ ($x, y$)] is a 2D vector in the r-space. The net effective potential in the
medium outside the Q-1DEG is expressed as
\begin{equation}
V(q,\omega; x)=\int dx'\, \epsilon^{-1}(q, \omega; x, x')\,V_{ex}(q, \omega ; x')\, .
\end{equation}
It is not difficult to demonstrate that the inverse dielectric function can be defined as follows.
\begin{equation}
\epsilon^{-1}(q, \omega ; x, x')=\delta (x-x') + \int dx'' \, V_{ee}(q, x-x'')\,\chi (q, \omega ; x'', x')\, .
\end{equation}
By making use of Eq. (61) in Eq. (63), we can write the total potential in the r-space in the form
\begin{align}
V({\boldsymbol r}, t)
&=\frac{1}{(2\pi)^2}\,\int dq \int d\omega\,e^{-i(\omega t-q y)}\,V(q, \omega; x)\nonumber\\
&=\frac{1}{(2\pi)^2}\,\int dq \int d\omega\,e^{-i(\omega t-q y)}\,
                  \int dx' \, \epsilon^{-1}(q, \omega; x, x')\, V_{ex}(q, \omega; x')\nonumber\\
&=\frac{e^2}{2\pi^2}\,\int dq \int d\omega\,e^{-i(\omega t-q y)}\, \int dx' \,
                \epsilon^{-1}(q, \omega; x, x')\nonumber\\
     & \hspace{3.5cm}  \times \int dt' \, e^{i[\omega t'-q y(t')]}\, K_0(q\mid x'-x(t')\mid)\, .
\end{align}
On substituting Eq. (65) into Eq. (58), one can write the induced force defined by
\begin{align}
{\boldsymbol F}=-\frac{e^2}{2\pi^2}\,\nabla \, \int dq \int d\omega\,e^{-i(\omega t-q y)}\, \int dx' \,
                \epsilon^{-1}(q, \omega; x, x')\nonumber\\
      \hspace{3.5cm}  \times \int dt' \, e^{i[\omega t'-q y(t')]}\, K_0(q\mid x'-x(t')\mid)
              {\Big |}_{{\boldsymbol r}={\boldsymbol r}(t)}\, .
\end{align}
On operating $\nabla$; limiting $x(t')=v_x t'+x_0$, $y=v_y t$, and $y(t')=v_y t'$; and decomposing the total
force into parallel [i.e., along y axis] and perpendicular [i.e., along x axis] components yields
\begin{align}
F_x=-\frac{e^2}{2\pi^2}\,\int dq \int d\omega\,e^{-i(\omega - q v_y)t}\, \int dx' \,
               \partial_x [\epsilon^{-1}(q, \omega; x, x')]\nonumber\\
      \hspace{3.5cm}  \times \int dt' \, e^{i(\omega - q v_y)t'}\, K_0(q\mid x'-v_x t'\mid)
              {\Big |}_{x=v_x t, x_0 =0}\, ,
\end{align}
where $\partial_x = \partial/{\partial x}$ and
\begin{align}
F_y=-\frac{i\,e^2}{2\pi^2}\,\int dq \int d\omega\,e^{-i(\omega - q v_y)t}\,
              \int dx' \, \epsilon^{-1}(q, \omega; x, x')\nonumber\\
      \hspace{3.5cm}  \times \int dt' \, e^{i(\omega - q v_y)t'}\, K_0(q\mid x'-x_0 \mid)
              {\Big |}_{x=x_0, v_x =0}\, .
\end{align}
Equations (67) and (68) are the exact results which we need to exploit further for the purpose of determining,
for instance, the rate of energy loss, the total energy loss, the stopping power, and/or the loss function
$P (q, \omega)$. Next, we specify a few configurations depending upon the spatial position of the incoming
fast-particle and the Q-1DEG [see, e.g., Fig. 3].

\begin{figure}[htbp]
\includegraphics*[width=8cm,height=8cm]{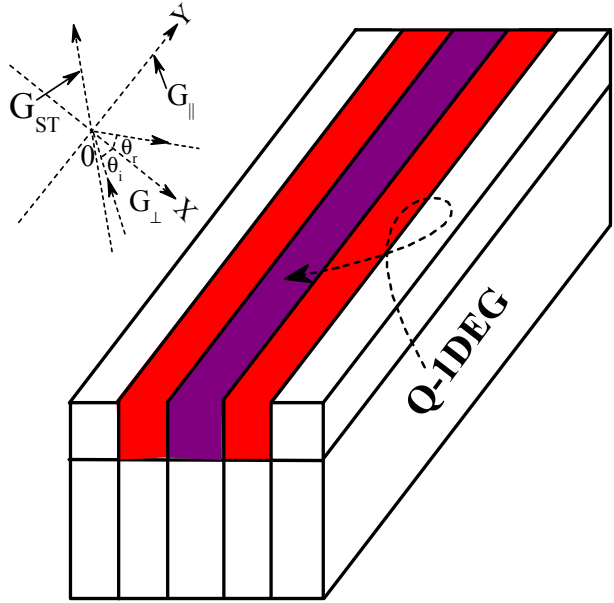}
\caption{(Color online) A schematic illustration of three principal configurations defining the geometry of the
electron beam with respect to the Q-1DEG held in the structure. The symbolic terms $G_{\|}$, $G_{\perp}$, and
$G_{ST}$ refer, respectively, to the parallel, perpendicular, and shooting-through configurations. The
$G_{\perp}$, strictly speaking, refers to the special case of the specular reflection [where the fast-particle
strikes the quantum wire and is back-scattered at $\theta_i=0^{\circ}=\theta_r$].}
\label{fig3}
\end{figure}

\paragraph{Parallel configuration}

This is a geometry where the coherent particle beam moves parallel to the y axis of the Q-1DEG at a distance $x=x_0$
from the center of the Cartesian coordinate system. Therefore, we would like to exploit Eq. (68) to write the rate of
energy loss first. Substituting Eq. (68) in Eq. (54) yields
\begin{align}
\frac{dW}{dt}
&=\frac{i\,e^2}{\pi}\,\int dq\, \int d\omega \,(q\, v_y)\, e^{-(\omega-q v_y)t}\, \delta (\omega-q v_y)\nonumber\\
 & \hspace{2.5cm}\times \int dx' \, \epsilon^{-1}(q, \omega; x_0, x')\, K_0(q\mid x'-x_0 \mid) \nonumber\\
&=\frac{i\,e^2}{\pi}\,\int dq\, \int dx' \,(q\, v_y)\, \epsilon^{-1}(q, \omega; x_0, x')\, K_0(q\mid x'-x_0 \mid)\, .
\end{align}
If we look at Eqs. (13) and (49), we can rewrite this equation in the form
\begin{equation}
\frac{dW}{dt}=\frac{i\,\epsilon_b}{2 \pi}\,\int dq\,(q\, v_y)\,V_s(q, \omega=q v_y; x_0, x_0)\, ,
\end{equation}
where $V_s (...)$ is the screened interaction potential of the two test electrons located at $x=x_0$ and $x'=x_0$.
Integrating both sides of the first equality of Eq. (69) with respect to time $t$ gives
\begin{align}
W&= 2\,e^2\,\int dq \int d\omega \, (q\, v_y)\, [\delta (\omega -qv_y)]^2 \nonumber\\
& \hspace{2.5cm} \times \int dx' \, {\rm Im}\big [\epsilon^{-1}(q, \omega; x_0, x')\big ]\,
                                         K_0 (q\mid x'-x_0 \mid)\nonumber\\
&=\int dq \int d\omega \, \hbar \omega \, P (q, \omega)\, ,
\end{align}
where the symbol $P (q, \omega)$ stands for the loss function [see above] and is defined as follows.
\begin{align}
P(q, \omega)= \frac{2e^2}{\hbar \omega}\, (qv_y)\, [\delta (\omega -qv_y)]^2
              \int dx' \, {\rm Im} \big [\epsilon^{-1}(q, \omega; x_0, x')\big ]\,\cdot K_0 (q\mid x'-x_0 \mid) .
\end{align}
This is the final expression for the loss function which should be dealt with at the computational level for the
parallel geometry. For any model employed to study specific elementary excitations, one needs to compute the
loss function $P(q, \omega)$ in order to obtain an explicit expression for the cross-section which is an
experimentally observable quantity.

\paragraph{Perpendicular configuration}

As the name suggests, this geometry refers to the situation where the fast-particle travels parallel to the x axis
and hits the Q-1DEG just perpendicularly. In other words, this geometry is a special case of the specular reflection
with the angles of incidence and reflection defined by $\theta_i=0^{\circ}=\theta_r$. Interestingly, it is necessary
and important to notice that this geometry requires particular attention regarding the sign of $v_x$: we clarify the
proper limits such that
\begin{align*}
x(t)=\left \{
\begin{array}{c}
+v_x t \,\,\,\,\, {\rm if \,\,\,\,\, t<0}\\
-v_x t \,\,\,\,\, {\rm if \,\,\,\,\, t>0}
\end{array}
\right .
\end{align*}
with $x_0=0$. For the purpose of deriving the rate of energy loss ($W'$), the total energy loss ($W$), or the finally
the loss function $P (q, \omega)$, we substitute Eq. (67) into Eq. (54) to write
\begin{align}
\frac{dW}{dt}=\frac{v_x e^2}{2\pi^2}\,\int dq \int d\omega\,e^{-i(\omega - q v_y)t}\, \int dx' \,
               \partial_x [\epsilon^{-1}(q, \omega; x, x')]\nonumber\\
      \hspace{3.5cm}  \times \int dt' \, e^{i(\omega - q v_y)t'}\, K_0(q\mid x'-v_x t'\mid)
              {\Big |}_{x=v_x t,}\, .
\end{align}
Integrating with respect to time gives the total energy loss defined by
\begin{align}
W&=\frac{v_x e^2}{2\pi^2}\,\int dq \int d\omega\,\int dx' \, \int dt\, e^{-i(\omega - q v_y)t}\,
               \partial_x [\epsilon^{-1}(q, \omega; x, x')]\nonumber\\
    &  \hspace{4.2cm}  \times \int dt' \, e^{i(\omega - q v_y)t'}\, K_0(q\mid x'-v_x t'\mid)
              {\Big |}_{x=v_x t}\nonumber\\
&=\int dq \int d\omega\, \hbar \omega \, P(q, \omega)\, ,
\end{align}
where the loss function $P(q, \omega)$ is given by
\begin{align}
P(q, \omega)=\frac{e^2}{2\pi^2 \hbar \omega}\,\int dx' \, &\int dt\, e^{-i\alpha t}\,
             v_x \, \partial_x [\epsilon^{-1}(q, \omega; x, x')]\nonumber\\
     \times &\int dt' \, e^{i\alpha t'}\, K_0(q\mid x'-v_x t'\mid)
              {\Big |}_{x=v_x t}
\end{align}
and the symbol $\alpha$ refers to the substitution $\alpha=\omega-q v_y$. Next, we solve very carefully the two
integrals with respect to time. We start with the first one to write
\begin{align}
I_1 &=\int^{\infty}_{-\infty} dt \, e^{-i\alpha t}\,v_x \partial_x [\epsilon^{-1}(x, x')]{\Big |}_{x=v_x t}\nonumber\\
&=\int^{\infty}_{-\infty} dx \, e^{-i\alpha x/v_x}\,\frac{\partial}{\partial_x} [\epsilon^{-1}(x, x')]\nonumber\\
&=\frac{i\alpha}{v_x}\,\int^{\infty}_{-\infty} dx \, e^{-i\alpha x/v_x}\,\epsilon^{-1}(x, x')\, .
\end{align}
We suppress the $(q, \omega)$ dependence of $\epsilon^{-1}(...)$ for brevity. Next, let us begin with the second
integral with respect to time $t'$ in Eq. (75) to write
\begin{align}
I_2 &=\int^{\infty}_{-\infty} dt \, e^{i\alpha t}\, K_0(q\mid x'-v_x t \mid){\Big |}_{x'=v_x t}\nonumber\\
&=\int^{0}_{-\infty} dt \, e^{i\alpha t}\, K_0(q\mid x'-v_x t \mid) +
   \int^{\infty}_{0} dt \, e^{i\alpha t}\, K_0(q\mid x'-v_x t \mid)\nonumber\\
&=\int^{\infty}_{0} dt \, e^{-i\alpha t}\, K_0(q\mid x'+v_x t \mid) +
   \int^{\infty}_{0} dt \, e^{i\alpha t}\, K_0(q\mid x'+v_x t \mid)\nonumber\\
&=2 \int^{\infty}_{0} dt \, \cos (\alpha t)\, K_0(q\mid x'+v_x t \mid){\Big |}_{x'=v_x t} \nonumber\\
&=\frac{2}{v_x}\, \int^{\infty}_{0} dy \, \cos (c y)\, K_0(q\mid x'+y \mid)\, ,
\end{align}
where $c=\alpha/v_x$ and $y$ is now any dummy variable. Obviously, the (suppressed) prime on $t$ does not matter.
The third equality takes care of the subtlety regarding the change of the sign of $v_x$ dictated by the first
(non-numbered) equation in here. Write [see, for example, \S 3.471.9 in Ref. 55]
\begin{equation}
\left .
\begin{split}
K_0 (q\mid x+y\mid)&=K_0 \Big (2\sqrt {q^2 (x^2+y^2+2 x y)/4} \Big )
\\
\text{and use}\,\,\,
K_0 \big (2\sqrt{a_1 a_2}\big )&=\int^{\infty}_{0} d\xi \,\, \frac{1}{2\xi}\,\, e^{-a_1\xi - a_2/\xi}
\\
\text{with}\,\,\,
a_2=1 \,\,\,\, {\rm and}&\,\,\,\, a_1=\frac{q^2}{4}\,\big (x^2+y^2+2 x y \big )
\end{split}
\right \}\, .
\end{equation}
With the aid of Eq. (78), Eq. (77) assumes the following form.
\begin{align}
I_2=\frac{1}{v_x}\,\int^{\infty}_{0} dy \,\cdot \cos (c y)\,
        \int^{\infty}_{0} d\xi \,\,\cdot \xi^{-1}\,\,\cdot e^{-(q^2/4) (x^2+y^2+2xy)\xi - 1/\xi}\, .
\end{align}
Rearranging the terms in here, one obtains
\begin{equation}
I_2=\frac{1}{v_x}\,\int^{\infty}_{0} d\xi \,\cdot\xi^{-1}\,\cdot e^{-(q^2 x^2/4)-1/\xi}\,\cdot I_{21}\, ,
\end{equation}
where [see, for example, \S 3.897.2 in Ref. 55]
\begin{align}
I_{21}&=\int^{\infty}_{0} dy \, \cdot \cos (c y)\,\cdot e^{-\beta y^2-\gamma y}\nonumber\\
&=\frac{1}{4}\,\sqrt{\pi/\beta}\,\cdot
\Big \{e^{(\gamma-i c)^2/(4\beta)}\,\cdot erfc\Big (\frac{\gamma-i c}{2\sqrt{\beta}}\Big )
      + e^{(\gamma+i c)^2/(4\beta)}\,\cdot erfc\Big (\frac{\gamma+i c}{2\sqrt{\beta}}\Big )\Big \}\, ,
\end{align}
where $\beta=q^2\xi/4$ and $\gamma=q^2 x \xi/2$. Substituting $I_{21}$ back in Eq. (80) enables us to write
\begin{align}
I_2=\frac{\sqrt{\pi}}{2 q v_x}\,\int d\xi \,\cdot \xi^{-3/2}\,\cdot e^{-(1+c^2/q^2)/\xi}\, \cdot D(q, c, x; \xi)
    \equiv F(q, \omega; v_x, v_y; x)\, ,
\end{align}
where the $F(...)$ has the units of time and the dimensionless symbol $D$($...$) is defined by
\begin{align}
D(q, c, x; \xi)=\Big [e^{-icx}\,\cdot erfc\Big (\frac{q^2 x \xi-2i c}{2 q \sqrt{\xi}}\Big )
      + e^{{+icx}}\,\cdot erfc\Big (\frac{q^2 x \xi+2i c}{2 q \sqrt{\xi}}\Big ) \Big ]\, .
\end{align}
Equation (75), with the aid of Eqs. (76) and (82), now assumes a simplified form
\begin{equation}
P(q, \omega)=\frac{i e^2}{2\pi^2 v_x}\,\frac{\alpha}{\hbar \omega}\,\int dx'\int dx''\, e^{-i\alpha x''/v_x}\,
     \cdot \epsilon^{-1}(q, \omega; x'', x')\,\cdot F(q, \omega; v_x, v_y; x')\, .
\end{equation}
This is the final form for the loss function which should be exploited at the computational level for the
perpendicular geometry. It is important to notice that while the electron spectroscopy employs a wavelength on
the order of the interparticle separation, the Raman spectroscopy exploits a radiation of longer wavelength.
Therefore it makes sense that through EELS -- in the impact regime -- one has (relatively) better chances to
obtain detailed structural information in the system.

\paragraph{Shooting-through configuration}

In order to understand this geometry, it is of vital importance to calculate the total work done by the induced
force comprised of both the parallel and the perpendicular components. As a result, we would be able to obtain
the total energy loss the fast-particle suffers from [in the process where the velocity component $\mid v_x\mid$
is constant $\Rightarrow$ particle shoots through]. We start with Eq. (54) together with Eq. (66) and integrate
over $t$ to write the total energy loss [with ${\boldsymbol r}={\boldsymbol r}(t)$ defined as follows:
$x= v_x t=x(t), y= v_y t=y(t)$]
\begin{align}
W=\frac{e^2}{2\pi^2}\,&\int dq \int d\omega \nonumber\\
\times &\Big [iq v_y \int dt \, e^{-i\alpha t} \int dx' \, \epsilon^{-1}(x, x')
            \int dt' \, e^{i \alpha t'}\cdot K_0(q\mid x'-v_x t' \mid) \nonumber\\
 & + v_x \int dt \, e^{-i\alpha t} \int dx' \, \partial_x [\epsilon^{-1}(x, x')]
            \int dt' \, e^{i \alpha t'}\cdot K_0(q\mid x'-v_x t' \mid) \Big ]{\Big |}_{x=v_x t}\, .
\end{align}
The ($q, \omega$) dependence of $\epsilon^{-1}(...)$ is suppressed for the sake of brevity. In the light of Eq. (76),
this can be cast in the following form.
\begin{align}
W=\frac{i e^2}{2\pi^2}\,&\int dq \int d\omega \nonumber\\
\times &\Big [\frac{q v_y}{v_x} \int dx' \int dx''\, e^{-i\alpha x''/v_x}\cdot\epsilon^{-1}(x'', x')
            \int dt' \, e^{i \alpha t'}\cdot K_0(q\mid x'-v_x t' \mid) \nonumber\\
 & + \frac{\alpha}{v_x} \int dx'\int dx'' \, e^{-i\alpha x''/v_x}\cdot\epsilon^{-1}(x'', x')
            \int dt' \, e^{i \alpha t'}\cdot K_0(q\mid x'-v_x t' \mid) \Big ]\, .
\end{align}
Summing up carefully the terms inside the square brackets piecewise leads us to write this as
\begin{align}
W&=\frac{i e^2}{2\pi^2}\,\frac{1}{v_x}\,\int dq \int d\omega \,\, \omega \nonumber\\
 &  \hspace{2.5cm}  \times \int dx' \int dx''\, e^{-i\alpha x''/v_x} \cdot \epsilon^{-1}(x'', x')\nonumber\\
 &  \hspace{3.5cm}  \times  \int dt' \, e^{i \alpha t'}\cdot K_0(q\mid x'-v_x t' \mid)\nonumber\\
&=\int dq \int d\omega \, \hbar \omega\, P(q, \omega)\, ,
\end{align}
where the loss function $P(q, \omega)$ is defined as
\begin{align}
P(q, \omega)=\frac{i e^2}{2\pi^2}\,\frac{1}{\hbar v_x}\,
          \int & dx' \int dx''\, e^{-i\alpha x''/v_x} \cdot \epsilon^{-1}(x'', x')\nonumber\\
 &  \times  \int dt' \, e^{i \alpha t'}\cdot K_0(q\mid x'-v_x t' \mid)\, .
\end{align}
It has become (nearly) customary for the scientists to depend so much on the computers that they want the latter
to do (almost) everything for them --  sometimes even at the expense of the desired accuracy. The software codes
available in some famous (and, of course, reliable) libraries [such as IMSL, NAG, ...etc.] can certainly do the
job, but the response may sometimes be ambiguous and hard to justify. This generally happens in the
circumstances involving, for example, the differentiation and/or integration of some subtle functions like
$K_0(z)\vert_{z \rightarrow 0}$. In the situations like this, it is always safer to do some analytical job and
feed only the indispensable stuff to the machine. As such, we realize that it is worthwhile to solve the last
integral in Eq. (88) and make the life better. We write
\begin{align}
I&=\int dt \, e^{i \alpha t}\cdot K_0(q\mid x-v_x t \mid)\nonumber\\
&=\frac{1}{v_x}\,\int dy\, e^{i c y}\cdot K_0(q\mid x-y\mid)\nonumber\\
&=\frac{1}{v_x}\,\int^{\infty}_{-\infty} dy\, e^{i c y}\, \int^{\infty}_{0} d\xi\,\, \frac{1}{2\xi}\,\,e^{-(q^2/4)(x^2+y^2+2xy)\xi-1/\xi}
\end{align}
In writing the last equality, we have made use of Eq. (78). To proceed further, we rearrange the terms in the
integrand and rewrite Eq. (89) such as
\begin{equation}
I=\frac{1}{2 v_x}\,\int^{\infty}_{-\infty} d\xi \,\, \xi^{-1}\,\cdot\,e^{-(q^2 x^2/4)\xi-1/\xi}\cdot
  \int^{\infty}_{-\infty} dy \,\, e^{-(q^2\xi/4)y^2+(q^2x\xi/2+ic)y}\, .
\end{equation}
In order to solve the second integral in this equation, we need to employ the identity [see, for example, \S 3.323.2
in Ref. 55]
\begin{equation}
\int^{\infty}_{-\infty}dy\,\, e^{-b_1 y^2 - b_2 y}=\sqrt {\frac{\pi}{b_1}} \cdot e^{b_2^2/(4 b_1)}\, ,
\end{equation}
with $b_1=q^2\xi/4$ and $b_2=-(q^2 x \xi/2+ic)$. This allows Eq. (90) to assume the simpler form
\begin{equation}
I=\frac{\sqrt{\pi}}{q v_x}\,e^{i c x}\int^{\infty}_{0} d\xi \,\,\xi^{-3/2}\,\cdot\,e^{-(1+c^2/q^2)/\xi}\, .
\end{equation}
In order to solve the integral in this equation, we make use of the identity [see, for example, \S3.325 in Ref. 55]
\begin{equation}
\int ^{\infty}_{0} dy \, y^{-3/2}\,\cdot e^{-c_1 y -c_2 /y}=\sqrt{\frac{\pi}{c_2}}\,\cdot e^{-2\sqrt{c_1 c_2}}
\end{equation}
with $c_1=0$ and $c_2=(1+c^2/q^2)$. Thus, Eq. (92) simplifies to the form
\begin{equation}
I=\frac{\sqrt{\pi}}{q v_x}\cdot e^{i c x}\cdot \frac{\sqrt{\pi}}{\sqrt{1+c^2/q^2}}
 =\frac{\pi}{\sqrt{\alpha^2 + q^2 v_x^2}}\cdot e^{i c x}\, .
\end{equation}
Therefore, Eq. (88), with the aid of Eq. (94), assumes the following form for the loss function
\begin{equation}
P(q, \omega)=\frac{i e^2}{2\pi}\,\frac{1}{\hbar v_x}\,\frac{1}{\sqrt{\alpha^2+q^2 v_x^2}}\,
             \int dx' \int dx''\, e^{i c (x'-x'')}\,\cdot \epsilon^{-1}(q, \omega; x'', x')\, .
\end{equation}
This is the final form for the loss function which should be treated at the computational level for the
shooting-through geometry. Interestingly, each of the three configurations envisioned here must, in
principle, be able to explore the electronic excitations in a Q-1DEG and provide us with an identical
answer. This is because the dominant contribution to the loss function ultimately comes from the
integration parts that involve the IDF, $\epsilon^{-1}(...)$. The prefactors do not matter much because
they can only influence the shape and/or size of the loss peaks, but not the peaks' location in energy.
The preference for a given configuration is then just a matter of taste. Equations (72), (84), and (95)
are thus the final results for the loss function in the respective geometries and, given the motivation
behind the present investigation, their derivation was considered to be essential.

\subsubsection{The inelastic light scattering}

The process of monochromatic light being scattered from a target inelastically -- where the scattered and
incident photons differ in energy (and hence in wavelength) -- is called inelastic light scattering, Raman
effect, Raman scattering, or Raman spectroscopy [after its discoverer  Sir C. V. Raman]. The course of
scattered photon being shifted to a lower (higher) frequency is termed as Stokes (antiStokes) shift. While
scattering from the acoustic (optical) phonons is called Brillouin (Raman) scattering, that from the
electronic excitations is preferably termed as inelastic light scattering in the literature. Raman
scattering is typically very weak and hence the intricacy with the Raman spectroscopy is to separate the
weak (inelastically) scattered signal from the strong (elastically) scattered signal. This is achieved by
using some sophisticated filters for laser rejection. The difference in energy between the incident and
scattered photons quantifies the (corresponding) excitations in the system.


\begin{figure}[htbp]
\includegraphics*[width=8cm,height=6.5cm]{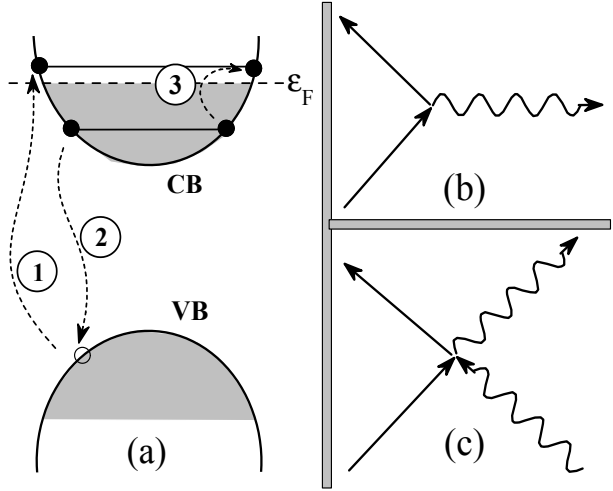}
\caption{(a) A schematic description of the difference between the resonant and the nonresonant processes
related with the inelastic light (or Raman) scattering. We sketch steps defining these processes by the
encircled numbers in the picture: step 1 refers to the case where the incident photon excites an electron
in the valence band (VB) into an excited state (above the Fermi level) in the conduction band (CB) leaving
a hole in the VB; step 2 refers to the case where an electron from the CB (below the Fermi level) recombines
with the hole in the VB emitting an outgoing photon shifted both in energy and momentum. Consequently, an
electronic excitation is created inside the CB through intermediate VB (IVB) states. This defines the
so-called resonant Raman scattering (RRS). The nonresonant Raman scattering (NRS), on the contrary, neglects
the IVB states and approximates the RRS process to take place {\em entirely within} the CB as depicted by
step 3 in the picture. To be succinct, NRS process is not equivalent to the RRS process: the latter involving
steps 1 and 2 depends explicitly on the incident photon energy, whereas the former referred to step 3 depends
only on the energy difference between the incident and scattered photons. This difference turns out to be
crucial in the RRS theory. In the right panel, (b) and (c) are, respectively, the Feynman diagrams for the
(light) scattering via ${\boldsymbol p}\cdot{\boldsymbol A}$ and ${\boldsymbol A}\cdot{\boldsymbol A}$ terms
in the Hamiltonian [47]. The solid (wiggly) line refers to the electron (photon) Green function.}
\label{fig4}
\end{figure}

Our concern here is the elementary electronic excitations such as single-particle and collective (plasmon)
excitations in the quantum wires (or Q-1DEG). These have been observed by several experimental techniques,
including the resonance Raman spectroscopy, which is, beyond doubt, the most versatile tool to observe the
low-energy plasmon excitations in the low-dimensional semiconducting nanostructures, including, e.g., the
quantum Hall systems subjected to strong applied magnetic fields.
A reader is referred to Fig. 4 for a brief introduction of the scattering process generally accounted for
in the ILS. We recall the differential scattering cross-section derived in Ref. 47 to write
\begin{eqnarray}
\frac{d^2 \sigma'}{d\omega d\Omega}
&=&-\,\frac{\hbar}{\pi}\, r^2_0\, \Big (\frac{\omega_s}{\omega_i}\Big )\,
      (\hat {e}_i \centerdot \hat{e}_s)^2\,[n_{B}+1]\,
                   {\rm Im}\big [\chi({\boldsymbol q}, \omega)\big ]\nonumber\\
&=&+\,r^2_0\, \Big (\frac{\omega_s}{\omega_i}\Big )\,
      (\hat {e}_i \centerdot \hat{e}_s)^2\,[n_{B}+1]\, S({\boldsymbol q}, \omega)\, .
\end{eqnarray}
where $n_{B}=[e^{\beta \hbar \omega}-1]^{-1}$, $r_0=e^2/m^* c^2$, $\omega_i (\omega_s)$, and $\hat{e_i}
(\hat{e_s})$ are, respectively, the well-known Bose-Einstein factor for photons, the classical electron
radius, the frequency of the incident (scattered) photon, and the unit polarization vector for incident
(scattered) photon in the Raman scattering process. The other symbols $\chi ({\boldsymbol q}, \omega)$
and $S({\boldsymbol q}, \omega)$ stand, respectively, for the interacting DDCF and the dynamical structure
factor (DSF) which are related through the familiar relation:
${\rm Im} \big [ \chi ({\boldsymbol q}, \omega)\big ]
= -(\pi/\hbar)\,\big [S({\boldsymbol q}, \omega) - S({\boldsymbol q}, -\omega)\big ]$. Some variants of
the scattering cross-section were derived in a different context (and with different methods) for the
conventional systems in the late sixties [56-58]. Remember, in deriving Eq. (96), we kept the formulation
for the ILS comply with the 3D systems [47], imposing no restriction whatsoever regarding the system's
dimensionality. Again, since the prefactors don't matter much, the imaginary part of the
DDCF ($Im [\chi (...)]$) or DSF ($S(...)$) is a direct measure of the scattering cross-section or the
Raman intensity in the scattering process of a given system. Obviously, the issue of the system's
dimensionality arises in relation with the derivation of $\chi (...)$ or $S(...)$. To that end, we begin
with expanding $\chi (q, \omega; x, x')$ in terms of the wave functions in the x direction such that
\begin{equation}
\chi (q, \omega; x, x')=\sum_{ijkl}\, \chi_{ijkl}(q, \omega)\,\phi^*_i(x)\,\phi_j(x)\,\phi^*_l(x')\,
\phi_k(x')\, .
\end{equation}
where the $\chi_{ijkl} (q, \omega)$ is defined by
\begin{equation}
\chi_{ijkl}(q, \omega)=\chi^0_{ij}(q, \omega)\,\delta_{ik}\,\delta_{jl} +
          \chi^0_{ij}(q, \omega)\,\sum_{mn}\,F_{ijmn}(q)\,\chi_{mnkl}(q, \omega)\, ,
\end{equation}
where $\chi^0_{ij}$ is the ($i, j$)th matrix element of the single-particle DDCF and is defined by
\begin{equation}
\chi^0 (q, \omega; x, x')=\sum_{ij}\, \chi^0_{ij}(q, \omega)\,\phi^*_i(x)\,\phi_j(x)\,\phi^*_j(x')\,
\phi_i(x')\, ,
\end{equation}
where
\begin{equation}
\chi^0_{ij}(q, \omega)=\Pi_{ij}(q, \omega)\, .
\end{equation}
Next, one can safely define
\begin{align}
\chi(q, \omega; q_x)
&=\int dx \int dx' \, e^{-i q_x (x-x')}\,\cdot \chi(q, \omega; x, x')\nonumber\\
&=\sum_{ijkl}\, \chi_{ijkl}(q, \omega)\cdot B_{ijkl}(q_x)\, ,
\end{align}
where
\begin{equation}
B_{ijkl}(q_x)=\int dx \int dx' \, \phi^*_i(x)\,\phi_j(x)\cdot e^{-i q_x (x-x')}\cdot \phi^*_l(x')\,
\phi_k(x')\, .
\end{equation}
Similarly, we can define
\begin{align}
\chi^0 (q, \omega; q_x)
&=\int dx \int dx' \, e^{-i q_x (x-x')}\,\cdot \chi^0 (q, \omega; x, x')\nonumber\\
&=\sum_{ij}\, \chi^0_{ij}(q, \omega)\cdot C_{ij}(q_x)\, ,
\end{align}
where
\begin{equation}
C_{ij}(q_x)=\int dx \int dx' \, \phi^*_i(x)\,\phi_j(x)\cdot e^{-i q_x (x-x')}\cdot \phi^*_j(x')\,
\phi_i(x')\, .
\end{equation}
The reader can have a legitimate concern as to why we have written the first equalities in Eqs. (101) and (103)
as if there prevails a translational invariance in the x direction, which is not the case. As explained in a
greater detail in Ref. 47, this is {\em mathematically} wrong. However, we do not have a better option. It is
because the intensity peaks in the Raman spectroscopy can only be interpreted in terms of the Fourier
transforms of the correlation functions. Despite all this fuss, the aforesaid analytical results are fairly
generic and unconstrained regarding the subband occupancy. However, we will specify them for practicality, e.g.,
in a two-subband model and determine the final expression for $\chi (...)$ useful for computing the Raman
intensity for the quantum wires [see Sec. II.D].

\subsection{The analytical diagnoses}

Ever since the man has learnt to grow and control their dimensionality, the low-dimensional semiconducting
systems have paved the way to some exotic (electronic) effects never before seen in the conventional host
materials [1]. The discovery of quantum Hall effects [both integral and fractional] -- that has changed our
basic notions of how an applied magnetic field at temperatures close to zero can give rise to some unparallel
quantum states -- is one of them. This tells us why the subsequent experiments on the low-dimensional systems
have generally been performed at lower temperatures. In order to turn the vast majority of our analytical
results into practicality, we intend to diagnose them at zero temperature, with restricted subband occupancy,
and simplified by the symmetry of the confining potential. The zero temperature limit also helps us simplify
the analytical results on the IDF and the DDCF used for investigating the IES and ILS.

\subsubsection{The zero temperature limit}

The zero-temperature limit has some very interesting consequences on the analytical results of a quantal
system: (a) this lets us replace the Fermi distribution function with the Heaviside unit step function,
i.e., $f(\epsilon)=\theta(\epsilon_F - \epsilon) =1$ (0) for $\epsilon_F >$ ($<$) $\epsilon$, where
$\epsilon_F$ is the Fermi energy in the system, (ii) this helps us convert the sum over ${k}$ to an
integral by using the summation replacement convention with respect to the 1D such as, e.g.,
$\sum_{k} \to [L/(2\pi)]\big [\int^{{k}_F}_{-{k}_F} d{k}\big ]$, and, most importantly,  (iii) this
permits us to calculate analytically the manageable forms of the polarizability function $\Pi_{nn'}(...)$.
Therefore, at $T=0$ K, $\Pi_{nn'}(...)$ in Eq. (10) takes the form
\begin{align}
\Pi_{nn'}(q, \omega)
&=\frac{2}{L}\,\sum_k \,\frac{f(\epsilon_{nk})-f(\epsilon_{n'k'})}
                             {\epsilon_{nk}-\epsilon_{n'k'}+\hbar\omega^+}\nonumber\\
&=\frac{m^*}{\pi q \hbar^2}\,
\ln \left [\frac{(\hbar \omega)^2 - (\epsilon_q + \Delta_{nn'} - \hbar q v_F)^2}
               {(\hbar \omega)^2 - (\epsilon_q + \Delta_{nn'} + \hbar q v_F)^2} \right ]
\end{align}
where $\epsilon_q = \hbar^2 q^2/2 m^*$, $\Delta_{nn'}=(n'-n) \hbar \omega^+$, $v_F=\hbar k_F/m^*$, and where
we assume that $n < n'$. Here $k_F$ and $v_F$ are, respectively, the Fermi wave vector and Fermi velocity in
the system. In the long wavelength limit (i.e., $q\to 0$), Eq. (105) yields the following:
$\Pi_{nn}\simeq \frac{n_{1D}\, q^2}{m^* \omega^2} + O (q^4)$ and
$\chi_{nn'} [=\Pi_{nn'}+\Pi_{n'n}]\simeq \frac{2\,n_{1D}\,\Delta_{nn'}}{(\hbar \omega)^2-\Delta^2} + O(q)$.
Here $n_{1D}$ is the 1D electron density in the $n$th subband. Notice that these long wavelength limits of
$\Pi_{nn'}(...)$ are independent of the system's dimensionality [1]. 

\subsubsection{Limiting the number of subbands}

As stated above, one must have the  matrix $\tilde{\epsilon} (q, \omega)$ of a finite order and recognize that,
theoretically, it is not feasible to calculate the excitation spectrum for a multiple subband model. This is
because the generalized dielectric-function matrix $\tilde{\epsilon}(...)$ happens to have the dimension of
$\eta^2 \times \eta^2$, where $\eta$ is the number of subbands in the model. It is therefore extremely hard to
handle enormous matrix of this order [for a very large $\eta$] and still enjoy the taste of the {\em analytical}
diagnosis. In addition, the literature seems to have no evidence whatsoever of any scientific innovation
emerging out of such inordinate complexity [1]. As such, we choose to keep the intricacy to a minimum and restrict
ourselves to a two-subband model [i.e., $n, n', m, m' \equiv 1, 2$] with only the lowest one occupied. This is quite
a sensible choice for such low-density, low-dimensional systems, particularly at low temperatures where most of the
(viable) experiments are performed. Consequently,  $\tilde{\epsilon}(q, \omega)$ is a $4 \times 4$ matrix defined by
\begin{equation}
\tilde{\epsilon}(q,\omega)=
\begin{bmatrix}
1-\Pi_{11}\,F_{1111} \ & \ -\Pi_{11}\,F_{1112} \ & \ -\Pi_{11}\,F_{1121} \ & \ -\Pi_{11}\,F_{1122} \\
-\Pi_{12}\,F_{1211} \ & \ 1-\Pi_{12}\,F_{1212} \ & \ -\Pi_{12}\,F_{1221} \ & \ -\Pi_{12}\,F_{1222} \\
-\Pi_{21}\,F_{2111} \ & \ -\Pi_{21}\,F_{2112} \ & \ 1-\Pi_{21}\,F_{2121} \ & \ -\Pi_{21}\,F_{2122} \\
-\Pi_{22}\,F_{2211} \ & \ -\Pi_{22}\,F_{2212} \ & \ -\Pi_{22}\,F_{2221} \ & \ 1-\Pi_{22}\,F_{2222}
\end{bmatrix}\, .
\end{equation}
Here $\Pi_{22}=0$ for the reason that the second subband is empty. Since the electronic excitations are computed
by employing the sustainability condition $\big | \tilde{\epsilon}(q,\omega) \big |=0$, Eq. (106) simplifies to
\begin{equation}
(1-\Pi_{11}\,F_{1111})\,(1-\chi_{12}\,F_{1212}) - \Pi_{11}\,\chi_{12}\,F^2_{1112}=0\, ,
\end{equation}
where $\chi_{12}[=\Pi_{12}+\Pi_{21}]$ is the intersubband polarizability function which accounts for both upward
and downward transitions. Notice that we may practice further simplification in solving Eq. (107) subject to the
nature of the confining potential [see in what follows].

\subsubsection{Symmetry of the confining potential}

It is becoming fairly known that, for a symmetric potential (as is the case here), the Fourier transform of the
Coulombic interaction $F_{ijkl}(q)$ is stringently zero for $i+j+k+l=$ an odd number [1]. The reason is that
the matching eigenfunction is either symmetric or antisymmetric under space reflection. This implies that the
intrasubband and intersubband excitations -- represented by the left and right parentheses in the (first) term in
Eq. (107) -- are now decoupled [because the term that enforces the coupling, $F_{1112}=0$]. Also, it is
significant to highlight that since the subindex $0$ ($1$) is permissible for the lowest (first excited) subband in the case of a confining harmonic potential,
Eq. (107) has to be conformed such that the subscript $1 \to 0$ and $2 \to 1$ for all practical
purposes. The ready-to-feed expressions for $F_{0000}(q)$, $F_{0101}(q)$, and $F_{0001}(q)$ [see Eq. (17)] are
given by
\begin{equation}
F_{0000}(q)=
\frac{2 e^2}{\pi \epsilon_b}\,\int dx\,\int dx'\, e^{-x^2}\, K_0({q\ell_c\mid x-x'\mid})
\, e^{-x'^{2}}\, ,
\end{equation}
\begin{equation}
F_{0101}(q)=
\frac{4 e^2}{\pi \epsilon_b}\,\int dx\,\int dx'\, x\, e^{-x^2}\, K_0({q\ell_c\mid x-x'\mid})
\, x'\, e^{-x'^{2}}\, ,
\end{equation}
and
\begin{equation}
F_{0001}(q)=
\frac{2\sqrt{2} e^2}{\pi \epsilon_b}\,\int dx\,\int dx'\,e^{-x^2}\, K_0({q\ell_c\mid x-x'\mid})
\, x'\, e^{-x'^{2}}\, .
\end{equation}
Here the symbols $x=z/\ell_c$ and $x'=z'/\ell_c$ are the dimensionless (dummy) variables. We illustrate
$F_{0000}(q)$, $F_{0101}(q)$, and $F_{0001}(q)$ versus the reduced momentum transfer $q/k_F$ in Fig. 5.
These plots clearly substantiate the aforesaid notion and hence lead us to infer that $F_{0001}(q)=0$ for
an arbitrary value of the propagation vector ($q$). It is not unexpected that $F_{0000}(q)$ turns out to be
predominant over the entire range of wave vector. This is clearly due to the fact that the charge density
at lower temperatures is largely concentrated in the ground state.

\begin{figure}[htbp]
\includegraphics*[width=8cm,height=9cm]{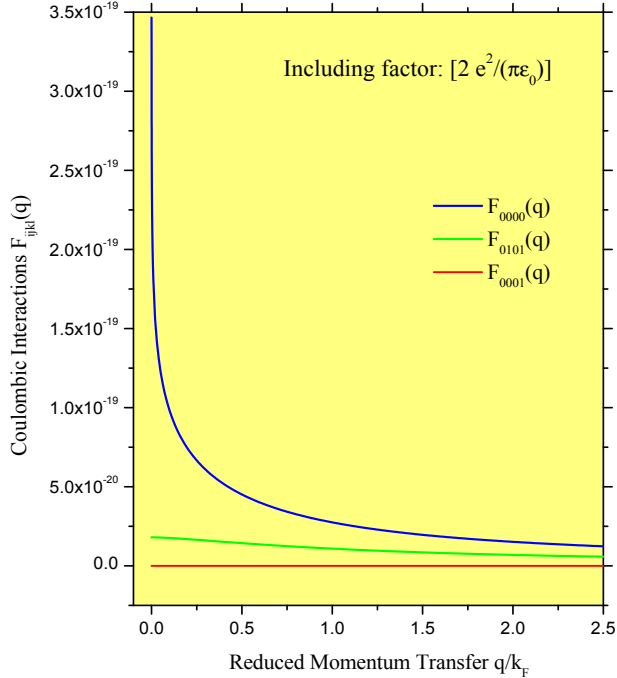}
\caption{The Fourier-transformed Coulombic interactions $F_{0000}(q)$, $F_{0101}(q)$, and $F_{0001}(q)$
plotted as a function of the reduced momentum transfer $q/k_F$. We call attention to the $F_{0001}(q)$
[in red] which is stringently zero over the entire range of the propagation vector $q$ [see the text].}
\label{fig5}
\end{figure}

\subsubsection{With respect to $\epsilon^{-1}(q, \omega; x, x')$ in the IES}

As we have seen in Sec. II.C, all the important results for the inelastic electron scattering are represented
in terms of ${\rm Im}[\epsilon^{-1}(...)]$. Therefore, it is considered to be interesting to simplify a few
relevant steps which remain quite involved in Sec. II.B. To that end, we need to exploit our strategy of
limiting to a two-subband approach. This implies that the composite index $\mu, \nu =nn'$ can take only three
values 11, 12, and 21. It suffices to say that $nn'$ cannot take the value 22 simply because the second subband
is empty. We recall Eq. (37) to stress that
$ \tilde{\Lambda}= \big [\tilde{I} - \tilde{\Pi}\tilde{\beta} \big ]^{-1}$.
Next, we diagnose the inverse dielectric function in Eq. (40). A thoughtful analysis leads us to write
\begin{equation}
\sum_{\mu\nu}\,L^*_{\mu}(x)\,\Pi_{\mu}\,\Lambda_{\mu\nu}\,S_{\nu}(x')=
L^*_{11}(x)\,P_{11}\,S_{11}(x') + L^*_{12}(x)\,P_{12}\,S_{12}(x')\, ,
\end{equation}
where
\begin{equation}
P_{11}=\frac{\Pi_{11}}{1-\Pi_{11}\,\beta_{1111}}\, ,
\end{equation}
and
\begin{equation}
P_{12}=\frac{\chi_{12}}{1-\chi_{12}\,\beta_{1212}}\, .
\end{equation}
As explained above, we still need to conform these equations such that the subscript $1 \to 0$ and $2 \to 1$
for every practical situation [47].  The aforesaid simplification turns out to be quite helpful in doing the
computation for the inelastic electron scattering, which depends so much on the use of the IDF.

\subsubsection{With respect to $\chi (q, \omega; q_x)$ in the ILS}

Here, we intend to seek a tractable expression for the interacting DDCF [$\chi ({\boldsymbol q}, \omega)=\chi
(q, \omega; q_x)$] involved in Eq. (96) and used to compute the (light scattering) differential
cross-section or simply the Raman intensity $I(\omega)={\rm Im}[\chi (q, \omega; q_x)]$ for the quantum wires.
This requires special attention to be paid to the analytical results discussed in Eqs. $(97)-(104)$ [see Sec.
II.C]. For a two-subband model, as is the case here, it turns out that $B_{1111}=C_{11}$ and $B_{1212}=C_{12}$.
Equation (97), for a two-subband model, gives rise to a $4\times 4$ matrix whose only nonvanishing elements
are 1,1; 2,2; 2,3; 3,2; and 3,3 -- the rest of them vanish for two obvious reasons: (i) the second subband is
unoccupied, and (ii) the confining (harmonic) potential is symmetric. It is not that difficult to determine
$\chi (q, \omega; q_x)$ which turns out to acquire a remarkably simple form. All we need to do is deal with
the aforesaid $4\times 4$ matrix and disentangle the correlation between its nonvanishing elements. A careful
analytical diagnosis yields
\begin{eqnarray}
\chi (q, \omega; q_x)
&=&\chi_{1111}\, B_{1111} + \big[\chi_{1212}+\chi_{1221}+\chi_{2112}+\chi_{2121}\big]\,B_{1212}\nonumber\\
&=&\frac{\chi^0_{11}}{1-\chi^0_{11}\,F_{1111}}\,B_{1111} +
   \frac{\chi^0_{12}+\chi^0_{21}}{1-\big (\chi^0_{12}+\chi^0_{21}\big )\,F_{1212}}\,B_{1212}
\end{eqnarray}
This is the final form of $\chi ({\boldsymbol q}, \omega)=\chi (q, \omega; q_x)$ to be exploited for studying
the light scattering cross-section expressed in Eq. (96). It is important to notice from Eq. (114) that (in
the two-subband model) the Raman scattering intensity $I(\omega)$ is the sum of two clearly distinguishable
parts: the intrasubband and intersubband collective (plasmon) excitations. Again, we do need to conform
Eq. (114) such that the subscript $1 \to 0$ and $2 \to 1$ everywhere.

\begin{figure}[htbp]
\includegraphics*[width=8cm,height=9cm]{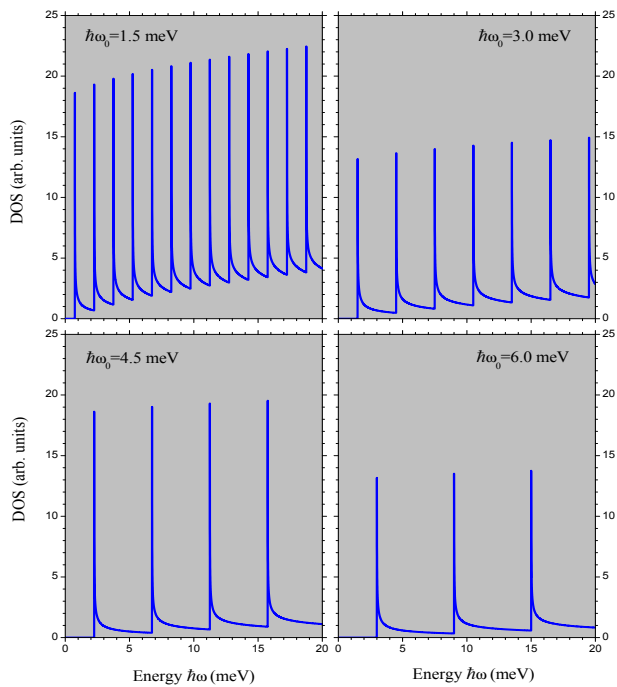}
\caption{(Color Online) The density of states versus the excitation energy. The number of DOS peaks reduces as
the confinement potential grows. This tendency is found to obey a simple mathematical rule [see the text]. The
heights of (some of) the peaks are rescaled for aesthetic.}
 \label{fig6}
\end{figure}

\subsubsection{The Density of states and the Fermi energy}

In condensed matter physics, (electronic) density of states (DOS) is one of the most fundamental characteristic
that defines and dictates the behavior of a system and plays key role in understanding its typical electronic,
optical, and transport phenomena. The same is necessarily true of the Fermi energy because all the transport
phenomena of a system are governed by the electron dynamics at/near the Fermi surface in the system. In the
normal physical situations, the DOS and the Fermi surface can (and generally do) influence each other in a very
intricate manner. Here we are interested in studying the DOS and the Fermi energy in a Q-1DEG in the absence of
an applied magnetic field. We start with Eq. (4) to derive the following expression for computing
self-consistently the density of states
\begin{equation}
D(\epsilon)=\frac{1}{\pi}\,\Big (\frac{2 m^*}{\hbar^2}\Big )^{1/2}\,
\sum_{n}\, \big (\epsilon - \epsilon_n \big )^{-1/2}\, \theta (\epsilon - \epsilon_n)
\end{equation}
and the Fermi energy
\begin{equation}
n_{1D}=\frac{2}{\pi}\,\Big (\frac{2 m^*}{\hbar^2}\Big )^{1/2}\,
\sum_{n}\, \big (\epsilon_F - \epsilon_n\big )^{1/2}\, \theta (\epsilon_F - \epsilon_n)
\end{equation}
where $\epsilon_n=(n+\frac{1}{2})\,\hbar\,\omega_0$, $k_F=(\pi/2)\,n_{1D}$ is the Fermi wave vector, and $n_{1D}$
the linear density (i.e., the number of electrons per unit length) of the system. It is (usually) customary to
subtract the zero-point energy $\epsilon_0$ ($=\frac{1}{2}\hbar\omega_0$) from the Fermi energy to compute the
effective Fermi energy of a system. Figure 6 shows the density of states versus the excitation energy for a given
value of the confinement potential. It is evident from all the four panels that the number of DOS peaks reduce as
the confinement potential grows. This tendency is found to obey a simple mathematical rule [1]:
$n_f={\rm Int}[n_i\times (\omega_i/\omega_f)]$, where $n_j$ is the number of peaks and $\omega_j$ is the
characteristic frequency; $j\equiv i, f$. To be specific, the subindex $i (f)$ refers to the initial (final) case
of the history. It must be emphasized that this (empirical) rule remains unfailingly true in the case of both zero
and non-zero magnetic field [1].

\begin{figure}[htbp]
\includegraphics*[width=8cm,height=9cm]{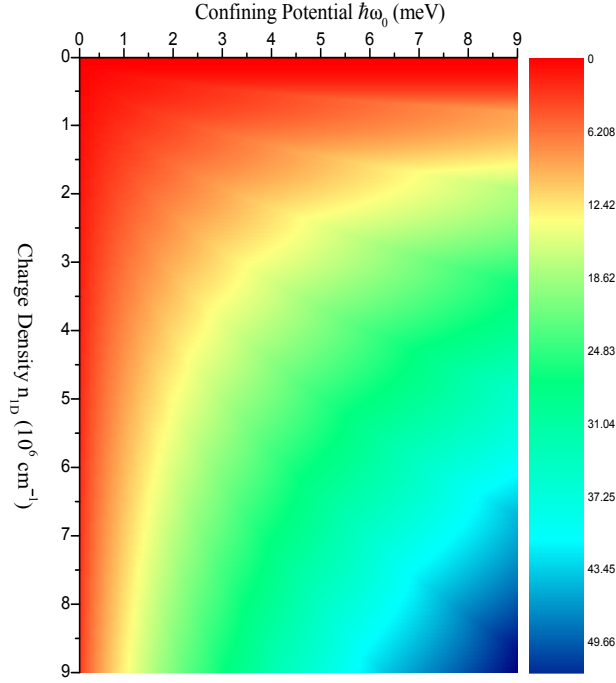}
\caption{(Color Online) A 3D plot of the (effective) Fermi energy [$\epsilon_{eff}=(\epsilon_F-\epsilon_0)]$ as
a function of the confinement potential $\hbar \omega_0$ and the charge-density $n_{1D}$ in the quantum wires.
The top of the structure seems to identify with the upper edge of a shell-cereal-bowl. The density of states
and the (effective) Fermi energy are seen to have an intricate relationship [see, e.g., Fig. 8].}
\label{fig7}
\end{figure}

Figure 7 depicts a 3D plot of the effective Fermi energy ($\epsilon_{eff}=\epsilon_F - \epsilon_0$) as a function
of the confining potential $\hbar \omega_0$, and the charge density $n_{1D}$ in the quantum wires. The top of the
structure apparently identifies with the upper edge of a shell-cereal-bowl. We have specifically considered the
most widely exploited GaAs/Ga$_{1-x}$Al$_{x}$As system. It has been known for quite some time that the density of
states for the quantum wires depends on the energy [with a power of $-0.5$] and that it takes on the spikes
separated by the magnitude of the confinement potential [1]. However, it does not seem to have been pointed out
before that the Fermi energy shows typical (periodic) dips which lie exactly midway between the consecutive peaks
of the DOS. Note that the larger the confinement potential energy ($\hbar\omega_0$), the smaller the number of
such dips. This is demonstrated in Fig. 8, which plots both the DOS and the Fermi energy on different scales for
specific values of the charge-density and the confinement potential.

\begin{figure}[htbp]
\includegraphics*[width=8cm,height=9cm]{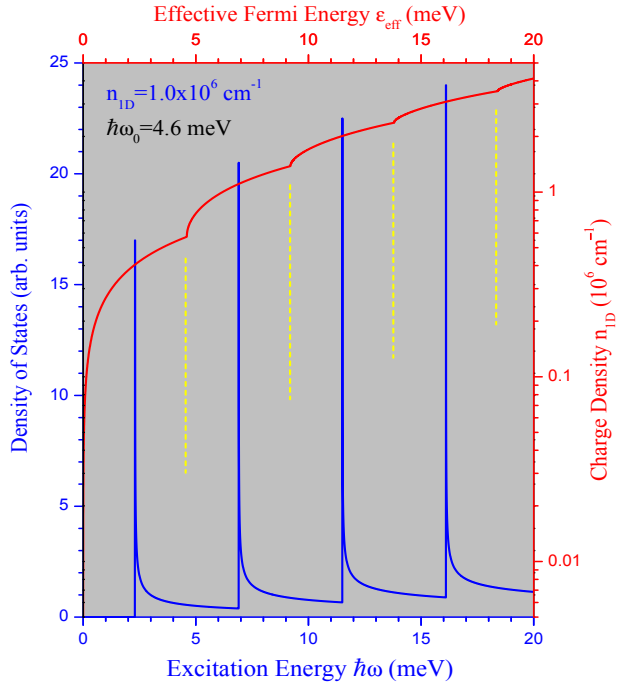}
\caption{(Color Online) The density of states versus the excitation energy (in Blue) and the effective Fermi
energy versus the charge density (in red). The broken (yellow) lines (only) indicate the dips (in the Fermi
energy) lying exactly in the center of the nearest DOS peaks.}
\label{fig8}
\end{figure}

\section{Illustrative Numerical Examples}

For the illustrative numerical examples, we focus on the isolated quantum wires fabricated out of the
GaAs/Ga$_{1-x}$Al$_{x}$As system. The material parameters used in the computation are: background dielectric
constant $\epsilon_{_b}=12.8$, effective mass $m^*=0.067 m_{_0}$, the subband spacing $\hbar\omega_{_0}=4.6$
meV, the (self-consistently determined) effective Fermi energy $\epsilon_{eff}=4.591$ meV for the linear
charge density $n_{1D}=5.72 \times10^{5}$ cm$^{-1}$, the effective confinement width of the harmonic
potential well, estimated as the FWHM from the extent of the Hermite function,
$w_{eff}=2\sqrt{2 \ln (2)}\sqrt{n+1/2}\,\ell_{_c}=26.182$ nm. Knowing $w_{eff}$ is quite important, because
it is this that helps define most of the limits of the integrals involved in the problem. It is
noteworthy that the Fermi energy $\epsilon_F$ varies with the variation in the charge density ($n_{1D}$)
and/or the confining potential ($\hbar \omega_{0}$). With all this input in hand, we aim at discussing
the excitation spectrum [made up of the single-particle and collective excitations], loss function for the
inelastic electron scattering, and the Raman intensity for the inelastic light scattering in a quasi-1DEG
within a two-subband model at T=0 K in the absence of an external magnetic field. The case of a nonzero
magnetic field is deferred to a future publication.

\begin{figure}[htbp]
\includegraphics*[width=8cm,height=9cm]{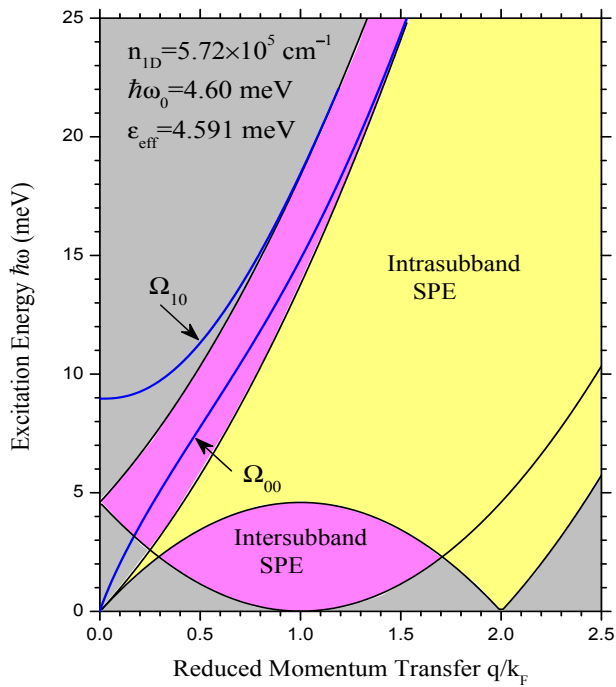}
\caption{The excitation spectrum of a quantum wire within a two-subband model: the energy ($\hbar \omega$) vs.
the reduced momentum transfer ($q/k_F$). The yellow (pink) shaded region refers to the intrasubband
(intersubband) SPE associated with the lowest (first excited) subband at $T=0$ K.
The bold lower (upper) curve (in blue) marked as $\Omega_{00}$ ($\Omega_{10}$) represents the intrasubband
(intersubband) CPE. After the CPE merge with the respective SPE, they become Landau-damped and cease to exist
as the bona-fide collective excitations with no life-time.}
\label{fig9}
\end{figure}

\subsection{Excitation spectrum}

Figure 9 illustrates the full excitation spectrum in a quasi-1DEG within a two-subband model in the framework
of Bohm-Pines' full RPA. We plot the excitation energy $\hbar\omega$ as a function of dimensionless momentum
transfer $q/k_F$. The excitation spectrum is, in general, made up of single-particle and collective (plasmon)
excitations. The yellow (pink) shaded region stands for the intrasubband (intersubband) single-particle
excitations (SPE) -- a continuum in the $\omega - q$ space where the polarizability function $\Pi (...)$ and
(consequently) the nonlocal, dynamic dielectric function $\epsilon (...)$ happen to have nonzero imaginary
parts. The bold lower (upper) curve (in blue) marked as $\Omega_{00}$ ($\Omega_{10}$) represents the
intrasubband (intersubband) collective (plasmon) excitations (CPE). Once the collective (plasmon) excitations
merge with (corresponding) single-particle excitations, they become Landau-damped [1], hence short-lived, and,
therefore, cease to behave like bonafide plasmon modes. A cursory glance at Fig. 9 leads one to infer that the
intrasubband collective excitation is longer-lived than its intersubband counterpart. While it is too much to
expect the analytical estimation of the critical point(s) associated with the propagation of the collective
excitations, it is not difficult to analyze: (i) why the lower branch of the intrasubband single-particle
continuum starts from the origin, attains a maximum at $q/k_F=1.0$, and then lowers to zero at$q/k_F=2.0$, (ii)
why the lower branch of the intersubband single-particle continuum exhibits its {\em minimum}
[{\em not zero}!] at $q/k_F=1.0$, and (iii) why the intersubband single-particle continuum starts at the
subband spacing at $q=0.0$.

There are two very subtile issues regarding the excitation spectrum shown in Fig. 9. First, even though a large
part of the intrasubband plasmon $\Omega_{00}$ propagates within the intersubband single-particle continuum,
the former does not bear the brunt of the latter [i.e., it does not suffer from the Landau damping]. And this
is because the intrasubband and intersubband excitations have different lineages that do not interbreed with
one another, rather than because they are decoupled due to the {\em symmetry} of the confinement potential.
Second, and the most important, issue here is the energy shift of the intersubband collective excitations from
the respective single-particle excitations in the long wavelength limit (i.e., $q\rightarrow 0$). This shift is
crucial and has reasonably been ascribed to the many-body effects such as depolarization shift [1], which is
relatively stronger in quantum wires than in quantum wells. In the lack of a desired quantal model, the
semi-classical estimate of the depolarization shift indicates the significance of the screening effects in the
quantum wires [1].

\subsection{Inelastic electron scattering}

In this section, we examine the computed loss functions $P (q, \omega)$ derived in Eqs. (72), (84), and (95),
respectively, for the parallel, perpendicular, and shooting-through configurations envisioned for the inelastic
electron scattering from the collective excitations in the Q-1DEG in a GaAs/Ga$_{1-x}$Al$_{x}$As system. As
stated above, we have neglected the prefactors (outside the signs of the respective integrals) for all the
illustrative examples presented here. This is because, as mentioned above, their inclusion can only cause an
insignificant change in the shape and/or size (but not in the energy position) of the loss peaks in the loss
spectrum. The other parameters unspecified in any (or all) of the three configurations are the same as those
used for computing the excitation spectrum in Fig. 9.

\begin{figure}[htbp]
\includegraphics*[width=8cm,height=9cm]{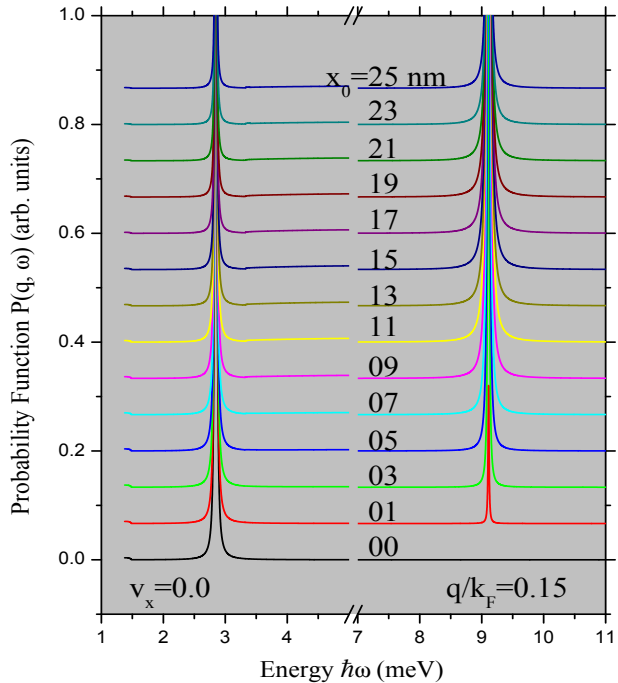}
\caption{(Color Online) The loss function $P (q, \omega)$ versus the energy $\hbar\omega$ for the parallel
configuration in a GaAs/Ga$_{1-x}$Al$_{x}$As quantum wire. Note that each curve corresponds to a different
value of $x_0$ -- the distance of the fast-particle from the origin. The parameters used are: the reduced
momentum transfer $q/k_F=0.15$ and the perpendicular component of the fast-particle velocity $v_x=0.0$.
Each curve is purposely displaced vertically for the sake of clarity. The y axis is rescaled with no loss
of generality. Notice the scale break on the x axis.}
\label{fig10}
\end{figure}

\subsubsection{The parallel configuration}

Figure 10 illustrates the loss function $P (q, \omega)$ against the energy $\hbar\omega$ for a fast-particle moving
parallel to y axis --i.e., the direction of the free motion -- over a Q-1DEG in the GaAs/Ga$_{1-x}$Al$_{x}$As
quantum wire. This case is comparatively special in the sense that the energy loss in the scattering process turns
out to be self-specified as $\hbar\omega=\hbar q v_y$. The noticeable features in Fig. 10 are the following. The
sharp $\delta$-like peaks at $\hbar\omega=2.8463$ meV and at $\hbar\omega=9.1103$ meV substantiate, respectively,
the intrasubband plasmon at $\hbar\omega=2.8468$ meV and the intersubband plasmon at $\hbar\omega=9.1110$ meV in
Fig. 9. This is considered to be an excellent agreement and hence justifies the use of the EELS to investigate
collective excitations in the quantum wires. The particle velocity of the lower (higher) loss peak is defined as $v_y/v_F=2.0666$ ($v_y/v_F=6.6146$). We have noticed that the larger the distance $x_0$, the smaller the amplitude
of the loss function $P (q, \omega)$. It is not unexpected that the energy position of the loss peaks remains
independent of the distance $x_0$. This makes sense because the momentum transfer $q/k_F$ is set
throughout. The other observations of the loss spectrum are: (i) only the particle velocities greater than the
Fermi velocity seem to be significant, and (ii) the larger the separation $x_0$, the smaller the rate of energy loss
($W'$), justifying the intuition. A unique characteristic of this configuration is that the fast-particle moving at
the distance $x_0=0$ does not seem to excite the intersubband collective excitation [see, e.g., the black
(lowest) curve in the picture]. This is understandable: the fast-particle (stringently) controlled to
pass through the center of the wire stays blind to the thickness of the quantum wire and hence remains ineffective
to excite the intersubband plasmon. Also, it is worth mentioning that if we zoom-in the spectrum in Fig. 10 --
before scaling the y axis -- we do see some weak signals indicating the likelihood of observing single-particle
excitations [see Fig. 9]. To sum up, Fig. 10 evidences the fact that the dominant contribution to the loss peaks
comes from the collective excitations.

\begin{figure}[htbp]
\includegraphics*[width=8cm,height=9cm]{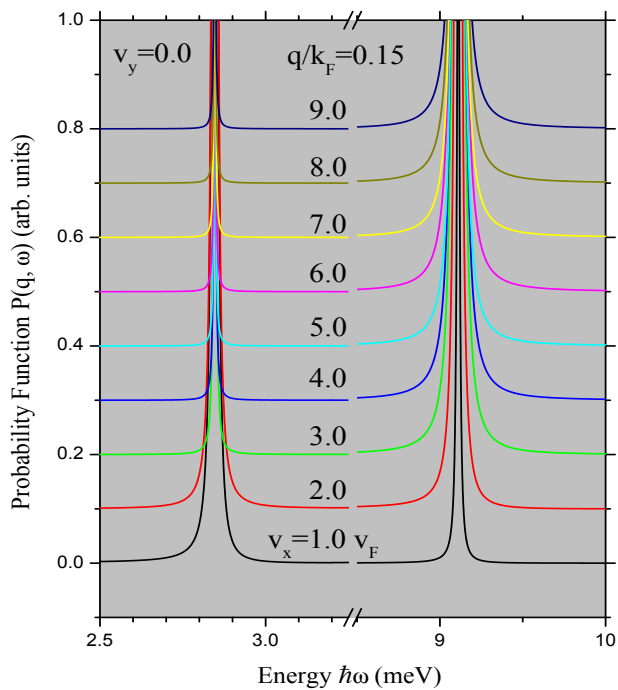}
\caption{(Color Online) The loss function $P (q, \omega)$ versus the energy $\hbar\omega$ for a fast-particle
specularly reflected from the Q-1DEG in a GaAs/Ga$_{1-x}$Al$_{x}$As quantum wire. Each curve corresponds
to a different value of the particle velocity $v_x$, which changes the sign at time $t=0$. The parameters
listed in the picture are: the reduced momentum transfer $q/k_F=0.15$ and the parallel component of the
fast-particle velocity $v_y=0$. The other remarks related with Fig. 10 still remain valid.}
\label{fig11}
\end{figure}

\subsubsection{The perpendicular configuration}

Figure 11 shows the loss function $P (q, \omega)$ plotted as a function of energy $\hbar\omega$ for a fast-particle
striking at and specularly reflected [with $\theta_i=0=\theta_r$ (see Fig. 3)] from the surface of a Q-1DEG in the GaAs/Ga$_{1-x}$Al$_{x}$As system. The sharp $\delta$-like loss peak at $\hbar\omega=2.8466$ meV
($\hbar\omega=9.1110$ meV) substantiates the intrasubband (intersubband) collective excitations at
$\hbar\omega=2.8468$ meV ($\hbar\omega=9.1110$ meV) for the corresponding value of the momentum transfer
$q/k_F=0.15$ in Fig. 9. Thus there is an outstanding agreement between the loss spectrum in Fig. 11 and the
excitation spectrum in Fig. 9. It is noteworthy that the total number of loss peaks must be less than or equal to
six, as is expected in a two-subband model for the Q-1DEG [see Fig. 9]. This is true
for all of the configurations employed to explore the IES. Also, it is noticeable that the sharp $\delta$-like
loss peaks (in energy) are independent of the variation in the particle velocity. This is obviously due to the fact
that the momentum transfer $q/k_F$ is kept constant throughout. Once more, the collective excitations are seen to be
the primary source of the loss mechanism. Intuitively, we believe that the perpendicular configuration favors a more
precise observation of the intersubband plasmons than any other arrangement.

\begin{figure}[htbp]
\includegraphics*[width=8cm,height=9cm]{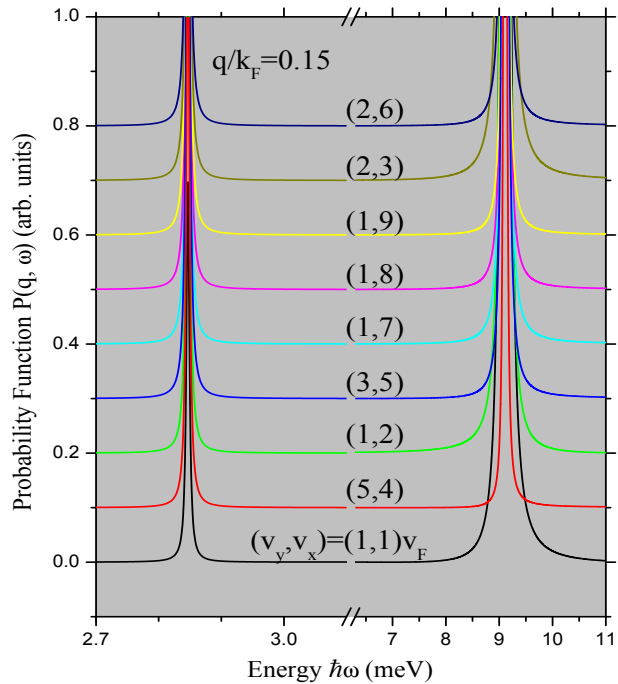}
\caption{(Color Online) The loss function $P (q, \omega)$ versus the energy $\hbar\omega$ for a fast-particle
shooting through a Q-1DEG in a GaAs/Ga$_{1-x}$Al$_{x}$As quantum wire. Each curve corresponds to a
different set of particle velocity ${\bf v}$ [$=(v_x, v_y)$]. The reduced momentum transfer $q/k_F=0.15$.
The other remarks related with Fig. 10 still remain valid. Rest of the parameters are the same as in Fig. 9}
\label{fig12}
\end{figure}

\subsubsection{The shooting-through configuration}

Figure 12 depicts the loss function $P (q, \omega)$ versus the energy $\hbar\omega$ for a fast-particle shooting
through a Q-1DEG in the GaAs/Ga$_{1-x}$Al$_{x}$As quantum wire. It is vital to notice that in this case we treat
the particle velocity ${\bf v}=$ as constant ($\Rightarrow$ particle shoots through). The sharp $\delta$-like
peaks at $\hbar\omega=2.8467$ meV and (relatively) broader peaks at $\hbar\omega=9.1109$ meV substantiate,
respectively, the intrasubband plasmon at $\hbar\omega=2.8468$ meV and the intersubband plasmon at
$\hbar\omega=9.1110$ meV, for the corresponding value of the momentum transfer $q/k_F=0.15$ in Fig. 9. This
ensures a very good agreement between the loss spectrum in Fig. 12 and the excitation spectrum in Fig. 9.
In addition, the (almost invisible) weak signals of smaller peaks (seen only after the zoom-in) let us believe in
the possibility of observing the single-particle excitations of Fig. 9. Despite the variation in the fast-particle
velocity, the energy-positions of both of the loss peaks remain intact. Just as in the previous two geometries, we
come to the conclusion that the the collective excitations remain conducive to the loss peaks in the spectrum. An
intriguing feature of this configuration, unlike the other two, is that the loss function does not abide by any
rule regarding the magnitude and/or the components of the particle velocity ${\bf v}$. The rest of the discussion
related to previous geometries is still valid.

While it is beyond the shadow of a doubt that the parallel and the perpendicular geometries are perfectly in the
reach of the current technology, one might feel a little bit skeptical about the shooting-through geometry (STG).
The root cause of this skepticism is likely the thought of how a fast-particle can shoot through a Q-1DEG embedded
in a host material. It seems as if the substrate materials cladding the Q-1DEG may obstruct the electron beam from
penetrating the whole system and into the detector. Given the reach of the current technology, the common-sense
belief is that a highly energetic, fast-particle must, in principle, overcome any such obstacle and shoot through
the whole system. Until the magical day arrives, the STG will remain, at least, of fundamental importance.

\begin{figure}[htbp]
\includegraphics*[width=8cm,height=9cm]{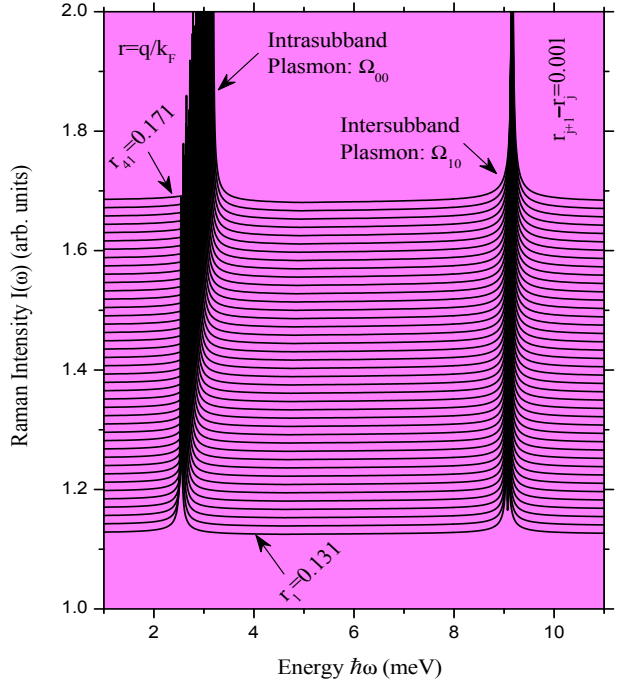}
\caption{The Raman intensity $I (\omega)$ vs. the energy $\hbar\omega$ for the inelastic light scattering
from a Q-1DEG in the GaAs/Ga$_{1-x}$Al$_{x}$As quantum wire. The long wavelength regime covers
$0.131 \le r=q/k_F \le 0.171$ and we feed $q_x=2 q$. The y axis is scaled with no loss of generality. Each
curve is displaced vertically for the sake of clarity. The rest of the parameters are the same as in Fig. 9.}
\label{fig13}
\end{figure}

\subsection{Inelastic light scattering}

In this section, we discuss the light scattering cross-section -- as defined in Eq. (96) -- which provides the
full (theoretical) interpretation of the inelastic light scattering experiments [or Raman spectroscopy] for a
given system. Since our interest lies in the quasi-1DEG, we worked out the relevant details in order to fully
determine the interacting DDCF [$\chi ({\boldsymbol q}, \omega)$, with ${\boldsymbol q}\equiv (q, q_x)$] via
Eqs. (97)-(104) in Sec. II. C. An explicit formal expression for $\chi ({\boldsymbol q}, \omega)$ was deduced
for a two-subband model at T=0 K in Eq. (114) in Sec. II.D. If we simply neglect the prefactors in Eq. (96) --
which only influence the shape and/or size of the spectral peaks but not their position -- the theoretical
efforts boil down to compute efficiently the imaginary part of the total DDCF [i.e.,
${\rm Im}[\chi ({\boldsymbol q}, \omega)]$], which is also proportional to the real part of the conductivity
in the light-absorption experiments. Note that different people use different terms to interpret the system's
response obtained through the computation of ${\rm Im}[\chi ({\boldsymbol q}, \omega)]$: some prefer the term
spectral weight and others the Raman intensity.

Figure 13 illustrates the Raman intensity $I (\omega)$ as a function of the excitation energy $\hbar\omega$ for
several values of the (parallel) momentum transfer ($q$) in the long wavelength limit specified by
$0.131 \le q/k_F \le 0.171$ and for a given value of the (normal) component of the momentum transfer $q_x=2 q $.
It is observed that there are only two prominent peaks in the energy range defined by
$1.0 \le \hbar\omega (\rm meV) \le 11$ for a given $q/k_F$. The lower and the higher Raman
peaks marked as $\Omega_{00}$ and $\Omega_{10}$ substantiate precisely the intrasubband and intersubband
collective (plasmon) excitations at the corresponding values of $q/k_F$ [cf. Fig. 9]. This justifies the
capability of the Raman spectroscopy as an experimental probe to investigate the collective excitations in the
quantum wires. It is worth mentioning, however, that we do not see, at least on this scale, any single-particle
Raman peaks corresponding to the intrasubband and/or intersubband single-particle continua in the excitation
spectrum in Fig. 9. But, if we zoom-in the spectrum in Fig. 13, with all the minute details to become
discernible, we do see some weak signals that (roughly) correspond to the single-particle excitations in Fig. 9.

\begin{figure}[htbp]
\includegraphics*[width=8cm,height=9cm]{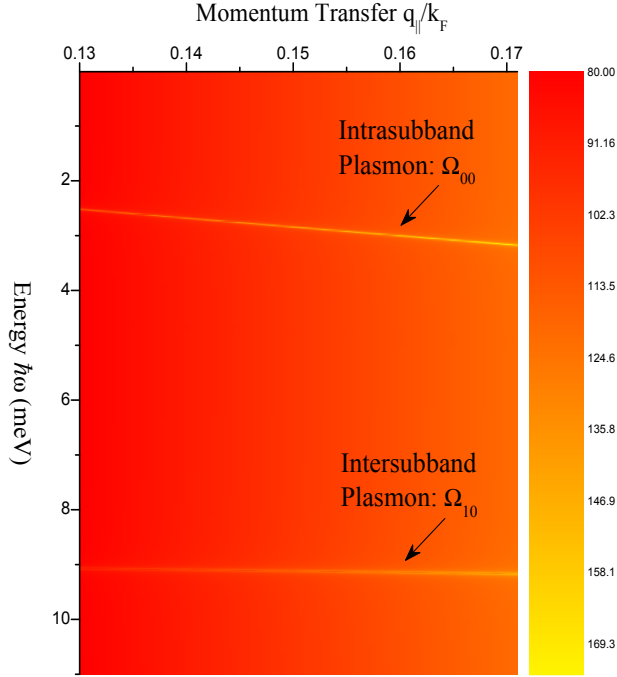}
\caption{The Raman intensity $I (\omega)$ vs. the excitation energy $\hbar\omega$ vs. the momentum transfer
$q/k_F$ for the inelastic light scattering from a Q-1DEG in the GaAs/Ga$_{1-x}$Al$_{x}$As quantum wires. We
feed $q_x=2 q$. This 3D figure is more instructive than its 2D counterpart [in Fig. 13].}
\label{fig14}
\end{figure}

Figure 14 shows a 3D plot of the Raman intensity $I (\omega)$, vs. the energy $\hbar\omega$, vs. the reduced
momentum transfer $q/k_F$. We focused on using (excessively) finest possible mesh of the momentum transfer
which makes the top axis of Fig. 14. The reason behind this choice was to make absolutely sure whether or not
there do exist any low-energy SPE-like peaks as observed, e.g., in the RRS experiments. We notice at
first hand that there are just the two clearly visible modes [marked as $\Omega_{00}$ and $\Omega_{10}$] which
substantiate the intrasubband and intersubband collective (plasmon) excitations [in Fig. 9], but no SPE-like
peaks -- at least not on this scale. While zooming-in the spectrum in Fig. 14 does reveal some weak signals
indicating the existence of the SPE-like peaks, those peaks have more to do with the SPEs in Fig. 9 and less
with those conjectured in the RRS experiments. We come to realize that this 3D figure is far more persuasive
and conclusive than its 2D counterpart [cf. Fig. 13].

The complexity regarding the low-energy SPE-like peaks in the RRS experiments and the lack of their theoretical
explanation is an old puzzle and is shared by all n-dimensional systems [$n\le 3$]. A few efforts -- embarking
on the (resonant) theory for the scattering cross-section [59] -- intended to give it a break, although garnered
some experimental support [60], yielded only a qualitative explanation.
The present (non-resonant) formalism for the scattering cross-section is not meant to shed light on the
sought-after clue to the famous RRS peaks. Let us keep an eye out on the El Dorado!

\subsection{On the inverse dielectric function}

The IDF can be utilized as an efficient alternative to compute the excitation spectrum as plotted in Fig. 9.
This is because the zeros of the dielectric function and poles of the IDF must produce exactly identical
results. The latter has certain advantage over the former, however. A thoughtful diagnosis of the IDF can
provide a substantive measure of the quantum transport in such nanostructures. To be specific, the imaginary
(real) part of the IDF determines the longitudinal (Hall) resistance $\rho_{yy}$ ($\rho_{xy}$) in the system.
Given the correlation between the total and single-particle DDCF
[$\chi (x, x')=\int dx''\, \epsilon^{-1} (x, x'') \chi^0(x'', x')$], the IDF can also furnish a significant
estimate of the Raman scattering cross-section. The quantity that directly impacts the quantum transport is
the spectral weight, ${\rm Im}$[$\epsilon^{-1}(q,\omega)$], which promises the collective (single-particle)
contributions at small (large) $q$.

\begin{figure}[htbp]
\includegraphics*[width=8cm,height=9cm]{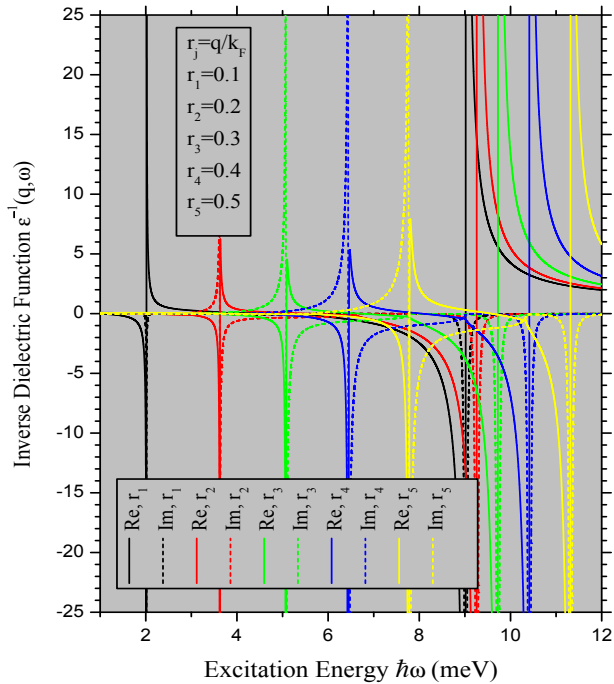}
\caption{(Color online) Inverse dielectric function $\epsilon^{-1}(q,\omega)$ vs. the excitation energy
$\hbar\omega$ for the given values of the momentum transfer $q/k_F$. The other parameters are the same
as in Fig. 9.}
\label{fig15}
\end{figure}

Figure 15 shows both the real and the imaginary parts of the IDF plotted as a function of the excitation energy,
for the given values of the momentum transfer $q/k_F$. For relatively small values of $q$ (as is the case here),
we expect the resonance peaks to identify only the collective (plasmon) excitations. We observe two sharp
resonances for each value of $q$ where the lower (upper) peak yields the intrasubband (intersubband) collective
excitations. To be specific, the resonance peaks occurring at $\hbar\omega=2.0156$, 3.6249,
5.0766, 6.4433, and 7.7698 reproduce the intrasubband plasmons and those at  $\hbar\omega=9.0171$, 9.2589, 9.7235,
10.412, and 11.315 yield the intersubband plasmons corresponding, respectively, to the values of 
$q/k_F=0.1$, 0.2, 0.3, 0.4, and 0.5. For higher values of $q/k_F$, one also starts discerning the single-particle excitations. For example, one can notice the crossing of the real and imaginary parts of $\epsilon^{-1} (...)$ at
$\hbar \omega=8.987$ and 10.312, respectively, for $q/k_F=0.4$ and 0.5. These values are the closest estimates of
the upper edge of the intersubband SPE in Fig. 9.

\begin{figure}[htbp]
\includegraphics*[width=8cm,height=9cm]{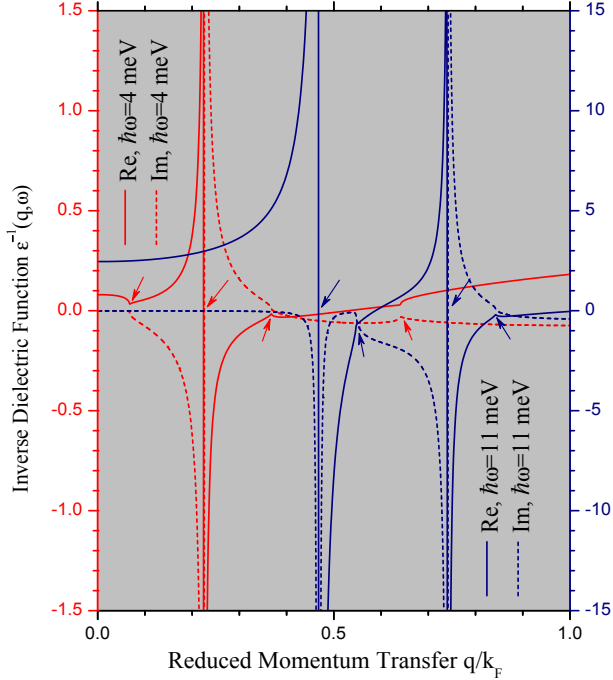}
\caption{(Color online) Inverse dielectric function $\epsilon^{-1}(q,\omega)$ vs. the momentum transfer
$q/k_F$ for the given values of the excitation energy $\hbar\omega$. The arrows indicate the substantive
values of $q/k_F$.}
\label{fig16}
\end{figure}

Figure 16 depicts the real and imaginary parts of the IDF plotted as a function of the momentum transfer, for
the given values of the excitation energy $\hbar\omega$. Again, we focus on the spectral weight
${\rm Im}$[$\epsilon^{-1}(q,\omega)$]. A careful look at Fig. 9 indicates that the sharp resonances for
$\hbar\omega=4.0$ (11.0) meV should yield collective (plasmon) excitations at $q/k_F=0.226$ (0.468 and 0.742).
This is exactly what we observe in Fig. 16. Similarly, the first, third, and fourth arrows
(in red) correspond to the intersubband and a pair of intrasubband single-particle excitations at
$q/k_F=0.068$ and (0.368 and 0.642), for $\hbar\omega=4.0$ meV. Likewise, the second and fourth arrows
(in blue) correspond to the intersubband and the intrasubband single-particle excitations at $q/k_F=0.548$ and
0.843, for $\hbar\omega=11.0$ meV [cf. Fig. 9]. Both Figs. 15 and 16 thus justify the notion that the IDF
represents an effective alternative for computing excitation spectrum of a quantum system.

\section{Concluding Remarks}

In summary, we have scrutinized the electron dynamics of Q-1DEG in a quantum wire within the framework of
Bohm-Pines' full RPA. Starting with the search for the exact (single-particle) eigenfunction and eigenenergy
of the Schr\"odinger equation for the 1D harmonic potential, we provide a methodical course to formulate the
generalized dielectric function, the inverse dielectric function, the dynamic screened potential, and the
Dyson equation correlating the interacting and the non-interacting DDCF. This provides us with a platform to
develop in a coherent manner the theory of inelastic electron scattering and of inelastic light scattering
in the quantum wires. Sec. II.D is an (unwonted) gift adding flavor to the wholesomeness of analytical
diagnoses. The strength of our efforts lies in this: {\em He who has a why to live can bear almost any how}.

The illustrative examples set out on the symmetry of the confining potential, the density of states, the Fermi
energy, the full excitation spectrum [comprised of both intrasubband and intersubband single-particle as well
as collective (plasmon) excitations], the loss spectra for the inelastic electron scattering, and the Raman
spectrum for the inelastic light scattering. However, we have attempted to be objective rather than expanding
on the illustrative (numerical) results. In other words, we have kept to the aim of providing thoroughly a
clear and concise string of tools to size up the research problem.

The interest in and importance of studying the inverse dielectric function (IDF) are manifold. The IDF not only
provides us with an effective alternative for studying the elementary electronic excitations, it also serves
the sole purpose for exploring the inelastic electron scattering. Moreover, the IDF is useful for investigating
not only the electronic and optical phenomena but also the transport ones. For instance, the imaginary (real)
part of the IDF sets to furnish a significant measure of the longitudinal (Hall) resistance in the system. Last
not least, the exact IDF as derived here knows no bounds with respect to the subband occupancy.

Interesting features worth adding to the problem are: the effects of (i) the coupling to the optical phonons,
(ii) an applied magnetic field, (iii) the spin-orbit interactions, (iv) the retardation through the
formulation in terms of the current-current correlation function, and (v) the many-body (exchange-correlation)
theory, to name a few. Even with the lack and limitation of the RRS experiments, it is worthwhile to include
the interband transitions in the general expression for Raman scattering cross-section. We feel emboldened to
state that the HREELS can be a potential alternative of the overexploited Raman spectroscopy for investigating
collective excitations in the quantum wires.

Currently, we have been generalizing the corresponding formalisms applicable to the quantum wells, quantum wires,
and quantum dots to include an applied (quantizing) magnetic field. The next step should be to incorporate the exchange-correlation effects, which share the inverse destiny with the systems of reduced dimensions:
their grip grows as the dimensions diminish.


\begin{acknowledgments}
The author feels grateful to Christian Sch\"uller and Sankar Das Sarma for very enlightening discussions and
communications with respect to the (resonant) Raman scattering in low-dimensional semiconducting nanostructures.
He has also enjoyed very stimulating communications with H. Ibach, and R.F. Egerton regarding the EELS. It is
a pleasure for him to thank Peter Nordlander, Naomi Halas, and Jan Kono for sharing their own experience in the
subject with him. He sincerely thanks Professor T.C. Killian for all the support and encouragement. Finally, he
wishes to express his deep appreciation to Kevin Singh for the unconditional help with the software during the
course of this investigation.
\end{acknowledgments}



\end{document}